\documentclass[11pt,a4paper]{article}
\pdfoutput=1

\usepackage{jheppub}
\usepackage{array}
\usepackage{xspace}
\usepackage[utf8]{inputenc}
\usepackage[T1]{fontenc}
\usepackage{afterpage}
\usepackage{rotating}

\newcommand{\cevns}
           {CE$\nu$NS\xspace{ }}

\makeatletter
\DeclareRobustCommand\recite[1]{\begingroup\@fileswfalse\cite{#1}\endgroup}
\makeatother

\graphicspath{{Figures/}}

\allowdisplaybreaks

\preprint{
  \begin{flushright}
    YITP-SB-2020-14\\
    IFT-UAM/CSIC-20-087\\
    IFIC-20-31
  \end{flushright}
}
  
\title{Determining the nuclear neutron distribution from Coherent Elastic neutrino-Nucleus Scattering: current results and future prospects}  

\author[a,b]{Pilar Coloma,}
\affiliation[a]{Instituto de Física Corpuscular, Universitat de
  València and CSIC, Edificio Institutos de Investigación, Calle
  Catedrático José Beltrán 2, E-46980 Valencia, Spain}
\affiliation[b]{Instituto de Física Te\'orica, Universidad Aut\'onoma de Madrid and CSIC, 
 Calle Nicol\'as Cabrera 13-15 Cantoblanco, 28049 Madrid, Spain}
\emailAdd{pilar.coloma@ift.csic.es}

\author[c]{Ivan Esteban,}
\affiliation[c]{Departament de Física Quàntica i Astrofísica and
  Institut de Ciències del Cosmos, Universitat de Barcelona, Diagonal
  647, E-08028 Barcelona, Spain}
\emailAdd{ivan.esteban@fqa.ub.edu}

\author[c,d,e]{M.~C.~Gonzalez-Garcia,}
\affiliation[d]{Institució Catalana de Recerca i Estudis Avançats
  (ICREA), Pg.\ Lluis Companys 23, E-08010 Barcelona, Spain}
\affiliation[e]{C.N.~Yang Institute for Theoretical Physics, Stony
  Brook University, Stony Brook, NY 11794-3840, USA}
\emailAdd{maria.gonzalez-garcia@stonybrook.edu}

\author[c]{Javier Menendez}
\emailAdd{menendez@fqa.ub.edu}

\abstract{
  Coherent elastic neutrino-nucleus scattering (CE$\nu$NS), a process
  recently measured for the first time at ORNL's Spallation Neutron
  Source, is directly sensitive to the weak form factor of the
  nucleus.  The European Spallation Source (ESS), presently under
  construction, will generate the most intense pulsed neutrino flux
  suitable for the detection of CE$\nu$NS.
  In this paper we quantify
  its potential to determine the root mean square radius of the point-neutron distribution, 
  for a variety of target nuclei and a suite of detectors.  To put our results in context we also
  derive, for the first time, a constraint on this parameter from the
  analysis of the energy and timing data of the CsI detector at the
  COHERENT experiment. }


\begin{document}

\maketitle
\section{Introduction}

In the Standard Model (SM), neutrinos can scatter off an atomic
nucleus via the weak neutral current (NC), through the exchange of a
$Z$ boson.  As long as the exchanged momentum $q$ remains
significantly smaller than the inverse of the nuclear size (which
typically requires $|q| \leq 50$~MeV for medium sized-nuclei), the
process can in principle take place coherently with the whole
nucleus. This translates into a drastic enhancement of the
cross-section for this type of neutrino interaction, which in the
coherent regime would be roughly proportional to the square of the
number of neutrons in the target nucleus, $N^2$.

The only observable signature for Coherent Elastic neutrino-Nucleus
Scattering (CE$\nu$NS) is a nuclear recoil with an energy between
sub-keV and a few keV (depending on the mass of the nucleus). Thus,
its detection presents a formidable task from the experimental point
of view.  Because of this it was not finally measured until 2017,
forty-three years following its theoretical description
\cite{freedman}, by the COHERENT collaboration~\cite{Akimov:2017ade}.
The experiment used the most intense neutron source in the world up
to date, provided by the Spallation Neutron Source (SNS) at Oak Ridge
National Laboratory (USA). Spallation sources are ideal to conduct an
experiment of this sort since neutrinos are produced from pion decay
at rest, which offers two main advantages: on one hand, the neutrino
spectra is well understood and can be computed analytically with high
precision; at the same time, the very low neutrino energies obtained
allow the coherence condition to be satisfied.

At COHERENT, the first measurement of this process was
obtained using a CsI[Na] detector of about
14.5~kg~\cite{Akimov:2017ade}, followed by a public data
release~\cite{Akimov:2018vzs}; a second measurement has been performed
with a Liquid Argon (LAr) detector~\cite{Akimov:2020pdx}, although the
data has not been made public yet.  These first results have already
triggered an intense activity in phenomenology, since the observed
CE$\nu$NS rates can be used to constrain both SM and Beyond the
Standard Model (BSM) physics scenarios. A non-exhaustive list of BSM
topics covered includes bounds on non-standard neutrino interactions
(NSI)
~\cite{nsi2,nsi1,nsi3,nsi4,nsi5,nsi6,nsi7,nsi8,nsi9,nsi10,Giunti:2019xpr,Denton:2018xmq,Coloma:2019mbs,Flores:2020lji},
constraints on neutrino electromagnetic properties
\cite{em1,em2,em3,em4,Papoulias:2019txv}, sterile neutrino
searches~\cite{ste1,carlos}, or searches for new weakly-interacting
particles from a hidden sector~\cite{dm1,dm2,dm3}.  On the other hand,
standard physics studies include new constraints on the weak mixing
angle \cite{wma1,wma2,wma3} at very low momentum transfer, as well as
studies of the nuclear structure factors of the target nuclei
~\cite{Cadeddu:2017etk,Ciuffoli:2018qem,Cadeddu:2018izq,Cadeddu:2019eta,Papoulias:2019lfi,Khan:2019cvi,Huang:2019ene,Canas:2019fjw,Cadeddu:2020lky,Miranda:2020tif}.

Since CE$\nu$NS is sensitive to the weak form factor, dominated by the
coupling to neutrons, this process can probe the distribution of
neutrons in nuclei.  This is precious information to complement proton
densities accessible with elastic electron
scattering~\cite{Donnelly:1984rg,Angeli:2013epw}.  At present the most
direct measurement of a neutron distribution comes from
parity-violating electron scattering in
$^{208}$Pb~\cite{Abrahamyan:2012gp}, also sensitive to the weak form
factor.  Alternative measurements rely on
nuclear~\cite{GarciaRecio:1991wk,Suzuki:1995yc,Clark:2002se,Trzcinska:2001sy,Lapoux:2016exf}
or electromagnetic~\cite{Tarbert:2013jze} reactions which probe both
neutron and proton distributions, but they lean on model-dependent
analyses (with uncertainties that are difficult to quantify).
Likewise, atomic parity violation experiments are also sensitive to the
nuclear neutron distribution, but they are subject to model-dependent
uncertainties from atomic many-body
calculations~\cite{Horowitz:1999fk,Brown:2008ib,Dzuba:2012kx,Viatkina:2019wsz}.

Therefore, CE$\nu$NS can shed light on neutron distributions in
nuclei, in particular their neutron radii.  The difference between the
radii of neutron and proton distributions is called neutron skin
thickness, or just neutron skin.  Its understanding impacts the limits
of existence~\cite{Erler2012} and size~\cite{Tanihata:2013jwa} of
atomic nuclei, and serves as an important test of first-principles
nuclear calculations~\cite{Hagen:2015yea,Lapoux:2016exf}.  Beyond the
structure of nuclei, the neutron skin can be related to the energy
needed to form isospin asymmetric nuclear matter---the symmetry
energy---and its variation with the nuclear
density~\cite{Brown:2000pd,Horowitz:2000xj,Centelles:2008vu,Tsang:2012se}. These
are key properties of the equation of state of neutron-rich matter,
which determines the size and structure of neutron
stars~\cite{Lattimer:2004pg,Lattimer:2006xb}.

The opportunity to complement and improve over the first
measurements of CE$\nu$NS using the upcoming European Spallation
Source (ESS) in Lund (Sweden) has been recently highlighted in
Ref.~\cite{Baxter:2019mcx}. The ESS will generate the most intense
neutron beams for multi-disciplinary science, and an order of
magnitude increase in neutrino flux with respect to the SNS. Using
novel detector technologies stemming from recent advances in dark
matter and neutrinoless double beta-decay experiments, the proposed
CE$\nu$NS@ESS would be able to maximally profit from the much higher
statistics available at the ESS.

Reference~\cite{Baxter:2019mcx} presented the physics potential to
constrain new physics scenarios in the neutrino sector using
CE$\nu$NS@ESS.  In the present work, on the other hand, the physics
scope is very different: we focus the study of nuclear structure with
the aim to determine the root mean square (rms) radius of the neutron
distribution ($R_n^{\rm pt}$) in the target nucleus, for several
detector materials (CsI, Xe, and Ge). In doing so the treatment of
systematic uncertainties is particularly relevant.
In this respect, technically, an
important difference with respect to Ref.~\cite{Baxter:2019mcx} is
that, in the present work, we also consider the effect of the energy
scale uncertainty, which, as we will show, has a non-negligible impact
on the determination of $R_n^{\rm pt}$.

In order to put our results into a larger context, we also derive the
present bounds on $R_n^{\rm pt}$ for CsI obtained from the analysis of
the COHERENT CsI data using both energy and timing information. While
other authors have studied the bounds on $R_n^{\rm pt}$
using energy
information alone, to our knowledge this is the first time that energy
and timing information are used to derive a bound on $R_n^{\rm pt}$.
As we will see, this brings additional synergies onto the
table and helps to improve the constraint with respect to the case
where only energy information is used when the analysis is performed
under the same assumptions. Furthermore, in our fit, we also implement
an improved background model and several choices of the quenching
factor (that gives the relationship between the number of
photoelectrons detected and the nuclear recoil energy of a given
event). In particular, following Ref.~\cite{Coloma:2019mbs} 
 we make use of the new parametrization of the TUNL data
which results on a new quenching factor function never considered
in the extraction of $R_n^{\rm pt}$ . We will show how the
inclusion of these effects leads to a spread in the extracted value
of $R_n^{\rm pt}$ beyond those previously considered in the literature.
Futhermore our results illustrate that current theoretical determinations
of the neutron radius may favour some values of the quenching factor. 

The article is structured as follows. In Sec.~\ref{sec:framework} we
introduce our notation and describe the methodology used in our
numerical fits and simulations. Our results are presented in
Sec.~\ref{sec:results}, both for current bounds obtained using the
COHERENT CsI data (Sec.~\ref{subsec:coh}) and for future sensitivities
expected at the ESS (Sec.~\ref{subsec:ESS}). Finally, we summarize and conclude in
Sec.~\ref{sec:conclusions}.

\section{Notation and framework}
\label{sec:framework}

The differential cross section for \cevns, for a neutrino with
incident energy $E_\nu$ on a nucleus of mass $M$, can be generically
written as \cite{Barranco:2005yy}
\begin{eqnarray}
  \frac{\mathrm{d}\sigma}{\mathrm{d}T} = \frac{G_F^2 M}{2 \pi}
&&  \left[ \left(F_V(Q^2)\,G_V + F_A(Q^2)\,G_A\right)^2 + \left(F_V(Q^2)\,G_V -
    F_A(Q^2)\,G_A\right)^2\left(1 - \frac{T}{E_\nu}\right)^2 \right.\nonumber\\
&&\left.
    - \left(F_V^2(Q^2)\,G_V^2 -F_A^2(Q^2)\,
  G_A^2\right) \frac{M T}{E_\nu^2} \right] \, ,
\label{eq:xsec}
\end{eqnarray}
where $G_F$ is the Fermi constant, $T$ is the recoil energy of the
nucleus (kinematically restricted to the interval $T\in \left[0, 2
  E_\nu^2/ (M + 2 E_\nu )\right]$) and $F_{V,A}$ are the vector and
axial form factors of the nucleus which are functions of 
the squared tree-momentum transfer $Q^2 \equiv |\vec{q}|^2 = 2 M T +
\mathcal{O}\left(\frac{T}{M}\right)$.  $G_V$ and $G_A$ are the
effective vector and axial couplings for the effective weak current of
the neutrino-nucleon interaction. The axial part of the cross section
is sensitive to the distribution of nucleon spins in the nucleus.
Because of the attractive nuclear pairing interaction, the sum over
nucleon spins cancels to a large extent~\cite{Klos:2013rwa}, so that
axial terms are typically smaller by a factor $1/N^2$.  Furthermore,
nuclei with even number of protons and neutrons have zero spin, so
that axial terms vanish.  With this, the differential cross section
for \cevns can be conveniently written as
\begin{equation}
\label{eq:xsec-short}
\frac{\mathrm{d}\sigma}{\mathrm{d}T} = \frac{G_F^2 M}{2 \pi} G_V^2
F_W^2(Q^2) \left[2 - \frac{M T}{E_\nu^2} - 2 \frac{T}{E_\nu} +
  \left(\frac{T}{E_\nu}\right)^2\right] \, ,
\end{equation}
where we have introduced the weak charge form factor of the nucleus $F_W \equiv F_V$, as is commonly done in the literature. 
For a nucleus with $Z$ protons and $N$ neutrons, $G_V$ 
can be written as a linear combination of the fundamental NC
couplings of the quarks and neutrinos:
\begin{eqnarray}
G_V & = & Z \left[2 \left(f^{u L} + f^{u R}\right) + \left(f^{d L} +
  f^{d R}\right) \right]+ N \left[\left(f^{u L} + f^{u R}\right) + 2
  \left(f^{d L} + f^{d R}\right) \right] \nonumber \\ & \equiv & Z\,
g_V^p+ N\, g_V^n \;, \label{eq:GV}
\end{eqnarray}
In these expressions, the left-handed neutrino NC couplings have
already been substituted into $G_V$, while the left- and
right-handed quark NC couplings $f^{qL}$ and $f^{qR}$ ($q = u, d$) as
well as the vector proton and neutron NC couplings $g_V^p$ and $g_V^n$ are
summarized in Tab.~\ref{tab:couplings} for convenience.

\begin{table}
\renewcommand{\arraystretch}{1.3}
\begin{center}
\begin{tabular}{c  c c }
\hline \hline 
  & SM (tree level) & SM expected value \\ \hline \hline
  $ f^{u L}$ & $ \frac{1}{2}-\frac{2}{3}\,s_W^2$ & 0.3458 \\ 
  $f^{u R}$ & $ -\frac{2}{3}\,s_W^2$ & -0.1552 \\ 
  $f^{d L}$ & $ -\frac{1}{2}+\frac{1}{3}\,s_W^2$ & -0.4288 \\ 
  $f^{d R}$ & $ \frac{1}{3}\,s_W^2 $ & 0.0777  \\ \hline
  $g_V^p$ & $\frac{1}{2} - 2\,s_W^2$ & 0.0301\\ 
  $g_V^n$ & $-\frac{1}{2}$ & -0.5116  \\\hline\hline
\end{tabular}
\caption{SM values of the up- ($u$) and down-quark ($d$) couplings to
  the $Z$ boson, for left- and right-handed particles ($L$ and $R$,
  respectively); as well as the vector couplings for protons ($p$) and
  neutrons ($n$). Here, $s^2_W \equiv \sin^2\theta_W$, where $\theta_W$
  is the weak mixing angle. The expressions on the left column
  correspond to the tree-level couplings in the SM. The values on the
  right column are taken from Ref.~\cite{Tanabashi:2018oca} and
  include propagator as well as vertex and box corrections, as
  detailed in Ref.~\cite{Erler:2013xha}. \label{tab:couplings} }
\end{center}
\end{table}

Neglecting relativistic Darwin-Foldy~\cite{Friar:1997js} and spin-orbit~\cite{Horowitz:2012we} corrections, typically below 0.1\% for the relevant $Q^2$ values in \cevns\cite{Horowitz:2012we,Horowitz:2012tj,Hoferichter:2016nvd,Hoferichter:2018acd},
the weak form factor of the nucleus can be written as
\begin{equation}
\begin{split}
F_W(Q^2) \simeq \frac{1}{G_V} & \left[ \left( g_V^p - g_V^p \frac{\langle r_p^2 \rangle}{6} Q^2 - g_V^n \frac{\langle r_n^2 \rangle}{6} Q^2 
 \right) \mathcal{F}^M_p(Q^2) + \right. \\
&\phantom{\bigg[} \left.
 \left( g_V^n - g_V^n \frac{\langle r_p^2 \rangle}{6} Q^2 - g_V^p \frac{\langle r_n^2 \rangle}{6} Q^2 
\right) \mathcal{F}^M_n (Q^2) \right] \, ,
\end{split}
\label{eq:FW}
\end{equation}
where $\langle r_p^2\rangle$ and $\langle r_n^2\rangle$ are the
squared charge radii for the proton and neutron,
respectively.
The terms in parentheses in Eq.~\eqref{eq:FW} represent the lowest-order
nucleon form factors, while $\mathcal{F}_{p,n}^M(Q^2)$ stand for the nuclear structure factors.
In particular, $\mathcal{F}_p^M(Q^2)$ and
$\mathcal{F}_n^M(Q^2)$ are the spin-independent proton and neutron
structure factors, respectively (see, e.g., Refs.~\cite{Hoferichter:2016nvd, Hoferichter:2018acd} for details). 
These encode the response of the
nucleus to the interaction, taking into account that the scattering
takes place with a complex many-body system. Their normalization is such that $\mathcal{F}_p^M(0) = Z$ and
$\mathcal{F}_n^M(0) = N$.

Several phenomenological parametrizations exist in the literature for
the nuclear structure factors, such as the Helm~\cite{Helm:1956zz},
symmetrized Fermi~\cite{Piekarewicz:2016vbn} and the
Klein-Nystrand~\cite{Klein:1999qj} parametrizations, among others.  In
our simulations, for concreteness, we use the Helm
parametrization~\cite{Helm:1956zz}
\begin{equation}
F_W(Q^2) = 3 \frac{j_1(Q R_0)}{Q R_0} e^{-Q^2 s^2/2} \, ,
\end{equation}
where $j_1$ is the first order spherical Bessel function, $R_0^2
\equiv \frac{5}{3} R_W^2 - 5 s^2$ and $s$ is set at $s = 0.9 \,
\mathrm{fm}$~\cite{Lewin:1995rx}.  Nevertheless, given the low
momentum transfers involved in CE$\nu$NS, it is enough to characterize
the structure factors using the first moment of the distribution in
$Q^2$.
For the nuclear structure factors ${\cal F}_i^M$  (where $i = p, n$)
this corresponds to the so-called point-proton and point-neutron distribution radii\footnote{
	Notice that point-proton and point-neutron radii are usually labeled in the literature 
	$R_{p,n}$. In here to avoid confusion with the notation of Ref.~\cite{Cadeddu:2020lky}
	we explicitly keep the index
	$^{\rm pt}$  when referring to the point-nucleon distribution radii.}
\cite{Ong:2010gf,Horowitz:2012tj,Horowitz:2012we,Cadeddu:2020lky}
\begin{equation}
\label{eq:Rpt}
  ({R^{\rm pt}_i})^2\equiv\left.-\,6\,\frac{1}{{\cal F}_i^M(0)}
  \frac{\partial
    {\cal F}_i^M(Q^2)}{\partial Q^2}\right|_{Q^2=0} \, .
\end{equation}
Likewise the weak form factor can be expressed in terms of the weak radius
$R_W$:
\begin{equation}
\label{eq:FW-RW}
    F_W(Q^2) \equiv F_W(0)\left[1 - \frac{1}{6} R_W^2 Q^2 \right] +
    \mathcal{O}(Q^4) \, .
\end{equation}
where, according to our normalization of the form factors,
$F_W(0)=1$. The weak radius can be then expressed in terms of $R^{\rm pt}_p$,
$R^{\rm pt}_n$, $\langle r_p^2\rangle$ and $\langle r_n^2\rangle$, as
\begin{equation}
\label{eq:RW-long}
\begin{split}
  R_W^2 = & \frac{Z}{G_V} \bigg( g_V^p (R^{\rm pt}_p)^2 + g_V^p \langle r_p^2
  \rangle + g_V^n \langle r_n^2 \rangle 
  \bigg) 
  +\frac{N}{G_V} \bigg( g_V^n (R^{\rm pt}_n)^2 + g_V^n \langle r_p^2
  \rangle + g_V^p \langle r_n^2 \rangle 
  \bigg).
\end{split}
\end{equation}

Equation~\eqref{eq:RW-long} can be simplified by
introducing the charge radius $R^2_\mathrm{ch}$, which is
precisely determined from elastic electron scattering.
Defined in terms of the charge form factor as in Eq.~\eqref{eq:Rpt}, with same approximations as Eq.~\eqref{eq:RW-long} the charge radius reads~\cite{Bertozzi:1972jff}
\begin{equation}
\label{eq:Rch}
R^2_\mathrm{ch} \simeq (R^{\rm pt}_p)^2 + \langle r_p^2 \rangle +
\frac{N}{Z}\langle r_n^2\rangle 
\, .
\end{equation}
This yields
\begin{eqnarray}
R^2_W &=& R_\mathrm{ch}^2+ \frac{N\,g_V^n}{G_V} \left[
  \bigg((R^{\rm pt}_n)^2-(R^{\rm pt}_p)^2\bigg)
+ \frac{Z^2-N^2}{Z\, N} \langle r_n^2\rangle\right] \, .
\label{eq:RW-short}
\end{eqnarray}
From Eq.~\eqref{eq:RW-short} we directly read that, in addition to the
information accessible in electromagnetic scattering experiments,
CE$\nu$NS provides independent information on the difference between
the rms radii of the neutron and the proton distributions, the neutron
skin.  Knowledge of the neutron skin is important to learn about
nuclear structure and test nuclear
models~\cite{Centelles:2008vu,Erler2012,Tanihata:2013jwa,Hagen:2015yea,Lapoux:2016exf,Cadeddu:2017etk}.
In addition, it can constrain the equation of state of neutron-rich
matter~\cite{Brown:2000pd,Horowitz:2000xj,Centelles:2008vu,Tsang:2012se},
a key ingredient for the structure of neutron
stars~\cite{Lattimer:2004pg,Lattimer:2006xb}.

From Eqs.~\eqref{eq:FW-RW} and \eqref{eq:RW-short} it is easy to see
that larger (smaller) values of $R^{\rm pt}_n$
tend to suppress (enhance)
the number of events with large momentum transfer. Since $Q^2 \simeq 2
M T $ this, in turn, affects the shape of the event
distribution as a function of the recoil energy of the nucleus, leading to a
similar suppression/enhancement of the tail.

In our simulations, we fit the data (either real data from COHERENT,
or simulated data for the ESS) to extract the weak radius, and use
Eq.~\eqref{eq:RW-short} to obtain the rms radius of the point-neutron
distribution
$R^{\rm pt}_n$. In doing
so, we use as inputs the tabulated nuclear charge radii
$R_\mathrm{ch}$ from Ref.~\cite{Angeli:2013epw}, together with the
Particle Data Group (PDG) values for the proton and neutron charge
radii\footnote{For the squared proton charge radius we have
  taken the {\sl small} values from muonic atom measurements, which seem in
  agreement with the results from modern elastic electron scattering
  experiments~\cite{Xiong:2019umf, Bezginov:2019mdi}, but differ from
  old electron scattering measurements by $\sim 5\%$.}: $\langle r_p^2\rangle=
0.70706 \, \mathrm{fm}^2$ and $\langle r_n^2\rangle\simeq -0.1161 \,
\mathrm{fm}^2$~\cite{Tanabashi:2018oca}. For convenience, the values 
of $R_{\rm ch}$ used in our calculations are summarized in Tab.~\ref{tab:Rch}. 
\begin{table}
\begin{center}
\begin{tabular}{|ccc|ccc|ccc|}
\hline\hline
\multicolumn{3}{|c}{CsI} & \multicolumn{3}{c}{Xe} & \multicolumn{3}{c|}{Ge} \\
\hline
\hline
Isotope & \% &$R_\mathrm{ch}$ (fm) & Isotope & \% &$R_\mathrm{ch}$ (fm) & Isotope & \% &$R_\mathrm{ch}$ (fm) \\
\hline
$^{133}\mathrm{Cs}$ & 50.0 & 4.80 & $^{132}\mathrm{Xe}$ & 26.9 & 4.79 & $^{74}\mathrm{Ge}$ & 36.7 & 4.07\\
$^{127}\mathrm{I}$ & 50.0 & 4.75 & $^{129}\mathrm{Xe}$ & 26.4 & 4.78 & $^{72}\mathrm{Ge}$ & 27.3 & 4.06\\
&&& $^{131}\mathrm{Xe}$ & 21.2 & 4.78 & $^{70}\mathrm{Ge}$ & 20.4 & 4.04\\
&&& $^{134}\mathrm{Xe}$ & 10.4 & 4.79 & $^{76}\mathrm{Ge}$ & 7.9 & 4.08\\
&&& $^{136}\mathrm{Xe}$ & 8.9 & 4.80 & $^{73}\mathrm{Ge}$ & 7.8 & 4.06\\
&&& $^{130}\mathrm{Xe}$ & 4.1 & 4.78 & & & \\
&&& $^{128}\mathrm{Xe}$ & 1.9 & 4.78 & & & \\\hline
\end{tabular}
\caption{Charge radius ($R_{\rm ch}$) and abundances (\%) used for the
  different isotopes considered in this work.  While for the CsI
  detectors we assume a 50\% number abundance of Cs and 50\% of I, for
  Xe and Ge we use their natural isotope abundances from
  Ref.~\cite{elementIsotopes}.  The values for the charge radii,
  taken from Ref.~\cite{Angeli:2013epw}, are measured
  to a few per mil precision, and therefore we ignore their error bars in our
  calculations. \label{tab:Rch}}
\end{center}
\end{table}
Finally note that, in the case of CsI, we assume the same
rms radius for the neutron distributions of Cs and I, because only one combined number can be extracted from \cevns.
This assumption seems reasonable given the sensitivity of current experiments and the similar values of $Z$ and $N$ for Cs and I.
\section{Results}
\label{sec:results}

\subsection{Signal and background event rates for a \cevns experiment}

At both the SNS and the ESS, the neutrino flux is predominantly
produced from pion decay at rest: while $\pi^-$ get rapidly absorbed
by the nuclei after being produced, the $\pi^+$ eventually decay at
rest into $\pi^ + \to \mu^+ \nu_\mu$. The prompt (monochromatic)
$\nu_\mu$ flux component is followed by a delayed contribution from
$\bar\nu_\mu$ and $\nu_e$, produced in the decay of the muon $\mu^+
\to \bar\nu_\mu \nu_e$.  The yearly energy spectrum of neutrinos
reaching a detector at a distance $\ell$ from the source, summed over
all neutrino flavours, reads
\begin{equation}
\label{eq:phi}
  \frac{\mathrm{d} \Phi_\nu}{\mathrm{d}E_\nu} (E_\nu) =
  N_\mathrm{POT}\times f_{\nu/p}\times \frac{1 }{4 \pi \ell^2}
  \left[\delta\left(E_\nu - \frac{m_\pi^2 - m_\mu^2}{2 m_\pi}\right) +
    \frac{16 E_\nu^2 (9 m_\mu - 16 E_\nu)}{m_\mu^4} \theta\left(E_\nu
    - \frac{m_\mu}{2}\right)\right] \, ,
\end{equation}
where $E_\nu$ is the neutrino energy, $m_\mu$ and $m_\pi$ are the muon
and charged pion masses, respectively, $N_{\rm POT}$ is the number of
protons on target (PoT) delivered per year, $f_{\nu / p}$ is the
neutrino yield per PoT (directly related to the pion yield per PoT), $\delta$ is the Dirac delta function
and $\theta$ is the Heaviside function.

At a \cevns experiment, three main sources of backgrounds should be
considered: (\textit{i}) steady-state  (SS) backgrounds (dominated by cosmic
ray interactions or by their by-products inside or in the surroundings
of a radio-clean detector); (\textit{ii}) beam-related backgrounds,
produced by neutrons escaping the target and reaching the detector;
and (\textit{iii}) neutrino-induced neutrons, that is, neutrons
produced in neutrino interactions inside (or in the surroundings of)
the detector. At COHERENT, the last background contribution was
determined to be very small \cite{Akimov:2017ade}, and therefore will
be also neglected here for the ESS. Beam-related background are also
expected to be sufficiently suppressed, see
Ref.~\cite{Akimov:2018vzs}.  While we include their expected
contribution in our fits to COHERENT, they will be neglected in our
ESS simulations for simplicity.  The most relevant background will therefore be the
SS contribution. In the case of COHERENT, this is estimated and
modeled using data taken when the proton pulse is turned off, as we
will discuss in more detail below.  At the ESS, the possible
contribution from this background can only be estimated from Monte
Carlo simulations, which however are not yet available. Therefore, we
conservatively assume it to be uniformly distributed in $T$, reading its
normalization from Tab.~1 in Ref.~\cite{Baxter:2019mcx}.

\subsection{Present bounds from COHERENT energy and timing data}
\label{subsec:coh}

For the sake of completeness and comparison with the existing
literature we start by performing an analysis of the present results
from COHERENT experiment using the detailed timing and energy
information of their data for CsI. In particular, we analyze the data
provided in Ref.~\cite{Akimov:2018ghi} for 308.1 live-days of neutrino
production, which corresponds to 7.48 GW-hr (or $\approx 1.76 \times
10^{23}$ protons on target). The events observed are binned in a
two-dimensional grid, using the number of photoelectrons observed
(equivalent to the nuclear recoil energy) and the time with respect to
the start of the beam pulse.  Although COHERENT data was used in
Refs.~\cite{Cadeddu:2017etk,Giunti:2019xpr} to extract the rms neutron radius, to our
knowledge this is the first time the timing information is included in
the determination of the neutron distribution radius in CsI. 
As we show below, this has an impact on the results.

We refer to the reader to Ref.~\cite{Coloma:2019mbs} for details on
our analysis of the COHERENT CsI data, while here we just summarize
some of the most relevant details of the fit.  In brief, we perform a
fit to the data following the COHERENT data release albeit with some
modifications. Namely, we study the dependence of the results with
respect to the choice of quenching factor (QF) as well as with the
implementation of the steady-state background component.  In
particular we perform the analysis for 3 choices of the QF and its
associated uncertainties: $(i)$ the energy-independent QF used in the
COHERENT data release; $(ii)$ the energy-dependent QF with reduced
uncertainties obtained in Ref.~\cite{Collar:2019ihs} (hereafter
referred to as ``Chicago QF''); and $(iii)$ a new QF parametrization
from our own fit to the calibration measurements performed by the TUNL
group~\cite{Akimov:2017ade, Akimov:2018vzs} (hereafter referred as
``Duke QF'').  In what respects to the background treatment, we use
two different parametrizations for the temporal behaviour of the SS
background: (a) we use the \emph{ad-hoc} exponential parametrization
prescribed in the experimental data release~\cite{Akimov:2018vzs}
(which however leads to an unexplained mild excess in the first two
time bins); and (b) we use our own parametrization of the temporal
behaviour of the background, based on a temporal fit to the number of
events detected when the neutrino beam is turned off.

\begin{figure}[t]
\centering
\includegraphics[width=\textwidth]{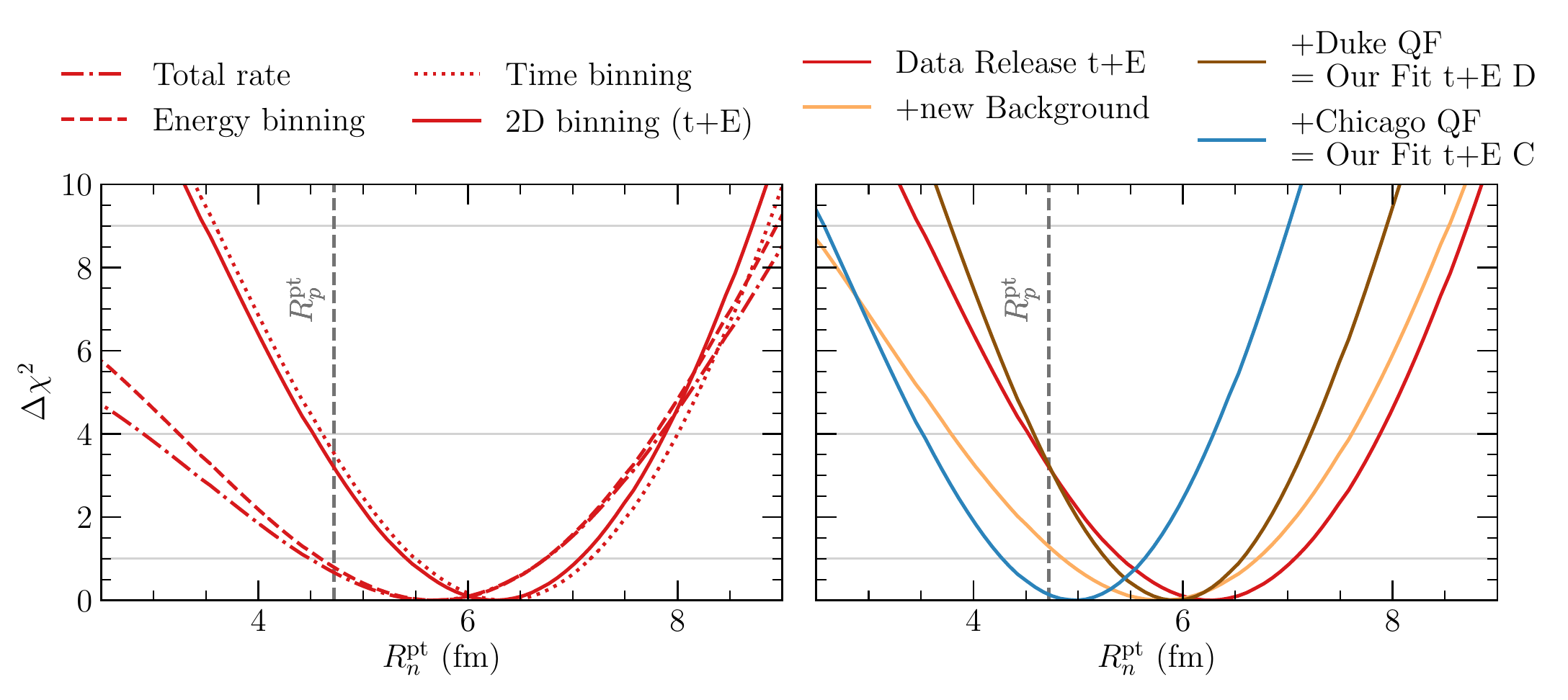}
\caption{
   $\Delta\chi^2$ as a function of the rms
  radius of the point-neutron distribution $R^{\rm pt}_n$ for Cs or I (assumed to be
  equal), for a variety of fits to COHERENT data as indicated by the labels. 
  In all cases shown in the left panel, the QF and
  background assumptions are those employed in the data
  release~\recite{Akimov:2018vzs}. On the right panel we show the
  dependence of the results on the assumptions for SS background
  modeling and QF implementation.  For convenience, the vertical line
  indicates the value of the rms radius for the proton
  distribution. This is taken as the average between their values for
  Cs ($R_p^{\rm pt,Cs} = 4.75$~fm) and I ($R_p^{\rm pt,I} = 4.70$~fm),
  obtained by substituting $R_{\rm ch}$ from
  Tab.~\ref{tab:Rch} into Eq.~\eqref{eq:Rch}. \label{fig:Rn-COH} 
}
\end{figure}

Our results for the fit are shown in Fig.~\ref{fig:Rn-COH}, where we plot the 
dependence of the $\chi^2$ on $R^{\rm pt}_n$ for the different models for the QF 
and background implementation used to fit the data. 
In the left panel we illustrate the dependence of our results on
$R^{\rm pt}_n$, for a fit which uses the same QF and SS background
parametrization as in the official data release.  The different lines
show the results obtained using only the total rate, energy
information, and/or timing information, as indicated by the labels.
Two salient features are identified right away from this panel.
First, comparing the dot-dashed and dotted lines, we find that the
inclusion of the energy dependence of the data does not lead to a
substantial improvement for the determination of $R^{\rm pt}_n$. This is so
because of the large systematic uncertainties assumed for the QF
employed in the data release, which do not allow to observe a
significant variation in the shape of the event distributions as $R^{\rm pt}_n$
is varied.  In contrast, once timing information is included the
$\chi^2$ increases significantly for small values of $R^{\rm pt}_n$, shifting
the best-fit towards slightly larger values.  This is mainly driven by
the small excess observed for the event rates in the first two time
bins (with respect to the SM prediction), which is in mild tension
with the prediction if we use an exponential fit to model the SS
background (as prescribed in the data release, see
Ref.~\cite{Coloma:2019mbs} for details). Therefore, the excess could
in principle be accommodated with a corresponding increase of the
prediction for the prompt neutrino contribution. Note that the prompt
neutrino flux is characterized by lower energies and therefore leads
to lower values of $T$, while the energies of the delayed component of
the flux are larger and contribute more to the high-energy tail of the
observed spectrum. Although the value of $R^{\rm pt}_n$ affects all neutrino
species (and therefore the prompt and delayed signals) in the same
way, a large value of $R^{\rm pt}_n$ would suppress the tail of the
distribution for large nuclear recoils; this, combined with an overall
increase of the total normalization (thanks to the large systematic
uncertainties affecting the fit) would effectively induce a change in
shape for the event distribution, mimicking an enhanced prompt
neutrino contribution.

On the other hand, the right panel of Fig.~\ref{fig:Rn-COH} quantifies
the effect on the fit due to changes in the QF and the background
treatment.  First, as forecasted in the above discussion, once the
\emph{ad-hoc} exponential parametrization of the temporal behaviour of
the SS background is substituted by a data-based modeling, the
prediction is able to fit better the observed data, and the induced excess 
at low time disappears.  As a result the best-fit is slightly
shifted to lower values, even when using the same QF implementation
and associated systematic error as in the data release. From the
comparison between the solid red (Data release t + E) and solid orange
(+ new Background) lines one can also see that the size of the
confidence regions is also slightly larger for the latter, as it is
often the case when the tension between data and background model is
eliminated.  Using our own fit to the TUNL data (Our fit t + E D)
results into a reduction of the uncertainty (as expected, since the
size of the systematic error associated to the QF is much smaller in
this case) but maintains the best-fit at the same value.  Instead, if
the Chicago QF
is used (Our fit t + E C), the allowed range of $R^{\rm pt}_n$
shifts to significantly lower values. The reason is that this QF leads
to a reduced number of predicted events (see Fig.~2 in
Ref.~\cite{Coloma:2019mbs}), which fall somewhat short to explain the
observed data.  In the fit, this can be partially compensated by
allowing for a smaller $R^{\rm pt}_n$, which leads to an enhancement
of the number of events.

\subsection{Future prospects at the ESS}
\label{subsec:ESS}

The ESS will soon generate the largest pulsed neutrino flux suitable
for the detection CE$\nu$NS.  In a recent 
article~\cite{Baxter:2019mcx} the potential for particle physics
phenomenology was quantified for the ESS, using a series of innovative
detector technologies specifically designed to detect very low nuclear
recoils, as those expected for \cevns. Here we summarize the main differences between 
the ESS and SNS neutrino sources, and describe how we adapt
the simulations performed in Ref.~\cite{Baxter:2019mcx} to the study
of the neutron distribution radius, while we refer the reader to
Ref.~\cite{Baxter:2019mcx} for additional details.

In what respects the assumed neutrino flux, the ESS is scheduled to
reach its design power of 5~MW by 2023.  This is to be compared to the
nominal 1~MW power of the SNS. Considering the higher proton energies
at the ESS (2~GeV, compared to 1~GeV at the SNS), this represents an
increase in average proton current at the ESS by a factor of 2.5, for
a total of $N_\mathrm{POT}= 2.8\times 10^{23}$ protons on target per
calendar year and approximately 5,000 hours of beam delivery per year.
An additional increase in the total neutrino flux arises from the
larger pion yield per proton at the ESS, due to the much higher proton
energies envisioned.  Based on a set of simulations performed in
Ref.~\cite{Baxter:2019mcx} we tentatively adopt a yield of $f_{\nu/p}=0.3$
neutrinos of each flavor ($\nu_{\mu},\bar{\nu}_{\mu},\nu_{e}$) per
proton for a ESS operating at 2~GeV, which is about 3.2 times larger
than the corresponding yield at the SNS.  Conversely, the time length
of the proton pulse at the ESS is much broader than that of the
SNS. This prevents the use of timing information for flavour
discrimination. Although this affects significantly the sensitivity to
some BSM scenarios, since the value of $R^{\text{pt}}_n$ affects the cross section
for all neutrino flavors in the same way, this will not be so relevant
for our physics case at hand. On the other hand, the much longer
proton pulse yields a duty factor of $4\times 10^{-2}$, almost two
orders of magnitude larger than the SNS duty factor of $6\times
10^{-4}$. Therefore, during the beam spill time window a larger number
of SS background events will enter the detector. Although these
backgrounds can be well-characterized using beam OFF data, their associated
statistical uncertainties have to be properly
accounted for in the simulations.

We compute the expected sensitivities for three different type of
detectors which use three different nuclear targets: (a) a cryogenic
undoped CsI scintillator array; (b) a high-pressure gaseous Xe
chamber; and (c) a low-threshold, multi-kg p-type point contact Ge
detector.  The rationale behind this choice is that the first CsI
detector allows for a direct comparison with the present bound from
COHERENT presented in the previous section, since both use the same
target material; while the other two detectors use a heavy (Xe)
and medium-mass (Ge) nuclear targets, allowing us to quantify the impact
on the sensitivity due to the choice of target nucleus. Finally, since
the optimal detector location for the ESS has not been identified yet,
in all our simulations we assume the detector distance to be $\ell=20$
m as a reasonable benchmark value.  In all cases we assume an exposure
of 3 years, corresponding to $8.4\times 10^{23}$ total PoT.

Table~\ref{tab:detectors} summarizes the main characteristics relevant
to the simulations, while we refer the interested reader to
Ref.~\cite{Baxter:2019mcx} for detailed descriptions and
characterizations of these detectors.

\begin{table}[t]
\newcolumntype{C}{ >{\centering\arraybackslash} m{1cm} }
\newcolumntype{D}{ >{\centering\arraybackslash} m{1.1cm} }
\newcolumntype{E}{ >{\centering\arraybackslash} m{1.7cm} }
\newcolumntype{F}{ >{\centering\arraybackslash} m{0.8cm} }
  \renewcommand{\arraystretch}{1.4} 
  \centering {
  \begin{tabular}{| >{\centering\arraybackslash} m{4cm} | C C D C E C F  | } 
      \hline \hline  
      Detector Technology
    & Target & Mass & SS bg. & $T_{th}$ & $\sigma_0$
    & $T_{\mathrm{max}}$ & $\mathrm{N}_\mathrm{ bins}$ \\[-0.2cm]
    & nucleus & (kg) & (ckkd) & (keV) &   (\%) &
    (keV) &\\ \hline\hline 
    Cryogenic scintillator & CsI & 22.5 & 10  & 1 & 30 & 46.1 & 21 \\ 
    High-pressure gas TPC & Xe & 20 & 10 & 0.9 & 40 & 45.6 & 17 \\ 
    p-type point contact Ge & Ge & 7 & 3 & 0.6 & 15 & 78.9 &42 \\ 
    \hline \hline
  \end{tabular}
}
	\caption{\label{tab:detectors} Summary of detector properties
          and background rates used in our sensitivity calculations.
          From left to right, the different columns indicate the target nucleus, total
          detector mass, steady-state (SS) background rates, nuclear recoil detection
          threshold in keV, energy resolution at
          threshold, maximum recoil energy considered, and the total
          number of bins used in the simulation (see text for
          details).  Backgrounds rates are listed in counts per keV, 
          kg and day (ckkd), before applying the
          $4\times10^{-2}$ reduction due to the ESS duty factor.  We
          conservatively adopt a flat SS background estimated at
          $T_{th}$ (where it is typically largest) for the whole energy
          range.  }
\end{table}

Our sensitivity calculations for the ESS use event distributions that
are binned in nuclear recoil energy. At a \cevns experiment, the
typical observable is usually the number of detected photoelectrons
(PE) in an event. The number of PE is related to the nuclear recoil by
the QF, which is determined from experimental calibration
measurements. For COHERENT, since the data are provided in terms of
the number of PE we have performed the analysis using that
variable. Conversely, since for the ESS we are dealing with a future
proposal, we have decided to bin the data in recoil energy instead. It
should be stressed out that, once the QF is known, both variables are
completely equivalent and the sensitivity analysis would give the same
results regardless of the variable used to bin the data. Also note
that, while we do not use a specific QF in our simulations for the
ESS, we \emph{do} consider an associated systematic error, as described
below.

Within each bin $i$ with reconstructed nuclear recoil energy
${T_\mathrm{rec} \in [T_i, T_{i+1}]}$, the expected number of events
per year $N_i$ can be computed as
\begin{equation}
N_i(R^{\rm pt}_n) = N^\mathrm{bkg}_i + \sum_j n_j^\mathrm{nuc}
\int_{T_i}^{T_{i+1}} \mathrm{d}T_\mathrm{rec} \int_0^\infty
\mathrm{d}T \mathrm{d}E_\nu \, R(T_\mathrm{rec}, T)
\frac{\mathrm{d}\sigma_j}{\mathrm{d}T} (E_\nu, T, R^{\rm pt}_n)
\frac{\mathrm{d}\Phi_\nu}{\mathrm{d}E_\nu} (E_\nu) \, ,
\label{eq:Nevts}
\end{equation}
where $N^\mathrm{bkg}_i$ the total number of background events per
year in that bin, $\frac{d\Phi}{dE_\nu}$ is the neutrino flux
integrated per year as in Eq.~\eqref{eq:phi}, and $n_j^\mathrm{nuc}$
is the number of target nuclei $j$ in the detector.
While for the CsI detector we assume a 50\% number abundance of
Cs and 50\% of I, for Xe and Ge we use their natural isotope
abundances as taken from Ref.~\cite{elementIsotopes}, see Tab.~\ref{tab:Rch}.

In Eq.~\eqref{eq:Nevts}, $R(T_\mathrm{rec}, T)$ is the the energy
reconstruction function. We assume it to be a Gaussian with a width
that depends on the recoil energy as: $\sigma(T) = \sigma_0
\sqrt{T/T_{th}}$, where $\sigma_0$ is the energy resolution at the
detection threshold ($T_{th}$), see Tab.~\ref{tab:detectors}.  For
each detector, the recoil energy bin sizes are chosen so that the
width of each bin is twice the energy resolution at its center.  We
consider all the kinematically available range (determined by the
condition $T \lesssim 2 E_\nu^2 / M $), for all detector
configurations.  For convenience, the total number of bins is provided
in the last column of the Tab.~\ref{tab:detectors}. 

In order to determine the sensitivity to $R^{\rm pt}_n$ we first simulate the
future {\sl data} as the expected number of events per bin,
$\bar{N}_i$, for a given detector and for an assumed true value for
the rms neutron radius, $R_{n}^\mathrm{pt, true}$. For sake of
concreteness, we assume $R_{n}^{\rm pt, true} =1.05 R_p^{\rm pt}$; however, we have
numerically checked that our results do not depend significantly on
the chosen value for $R_{n}^\mathrm{pt, true}$ as long as it lies within the
range $\sim \pm 25 \% \, R_p^{\rm pt}$.  Once the event distribution for the
\emph{data} ($\bar{N}_i$) has been simulated, and the expected event
rates ($N_i$) have been computed as a function of $R^{\rm pt}_n$, a binned
$\chi^2$ is built.  Systematic uncertainties are implemented using the
pull method. In doing this, a set of nuisance parameters ${\eta_j}$ is
included in the fit, each of them with an associated prior
$\sigma_j$. A penalty term is then added to the $\chi^2$ for each
source of systematic errors, and the minimum of the $\chi^2$ is
determined after marginalization over all the nuisance parameters
included in the fit.  We consider three different types of systematic
uncertainties:
\begin{enumerate}
\item An error on the total signal normalization. To implement it, we
  substitute $N_\mathrm{POT} \rightarrow N_\mathrm{POT} (1 +
  \eta_\mathrm{norm} \sigma_\mathrm{norm}) $ in Eq.~\eqref{eq:Nevts}.
  We set $\sigma_\mathrm{norm} = 0.1$~\cite{Baxter:2019mcx}.
\item An error on the total background normalization. To implement it,
  we substitute $N^\mathrm{bkg}_i \rightarrow N^\mathrm{bkg}_i (1 +
  \eta_\mathrm{bkg} \sigma_\mathrm{bkg})$ in Eq.~\eqref{eq:Nevts}.  We
  set $\sigma_\mathrm{bkg} = 0.05$~\cite{Baxter:2019mcx}.
\item A systematic error affecting the energy scale (ES), directly
  related to the QF uncertainty. To implement it, we substitute
  $T_\mathrm{rec} \rightarrow T_\mathrm{rec} (1 +
  \eta_\mathrm{ES}\,\sigma_\mathrm{ES}(T_{\rm rec}))$ in
  Eq.~\eqref{eq:Nevts}.  Here, $\sigma_\mathrm{ES}$ is the prior ES
  uncertainty, which may depend on the reconstructed recoil energy.
  Since we find that this uncertainty is the one that has the largest
  impact on our results (see below for details), we will consider three different assumptions
  for $\sigma_\mathrm{ES}$: negligible; an energy-independent 5\%
  uncertainty; and an energy-dependent uncertainty which decreases
  linearly from 5\% at $T_\mathrm{rec, th}$ to 1\% at
  $T_\mathrm{rec,max}$.
\end{enumerate}
With all these ingredients, our $\chi^2$ function reads
\begin{equation}
  \chi^2(R^{\rm pt}_n) = \min_{\left\{\eta\right\}  }  \left[\sum_i 2 \left( N_i(R^{\rm pt}_n, \left\{\eta\right\}) -
    \bar{N}_i + \bar{N}_i \log\frac{\bar{N}_i}{N_i(R^{\rm pt}_n, \left\{\eta\right\} )}\right)
    + \eta_\mathrm{norm}^2 + \eta_\mathrm{bkg}^2 + \eta_\mathrm{ES}^2
    \right] \, .
\end{equation}
where $\bar{N}_i$ are the assumed {\sl data} event rates at face
value, i.e., $\bar{N}_i= N_i(R^{\rm pt}_n = R_n^\mathrm{pt,true}, {\eta_j}=0)$.

\begin{figure}[t]
\centering \includegraphics[width=\textwidth]{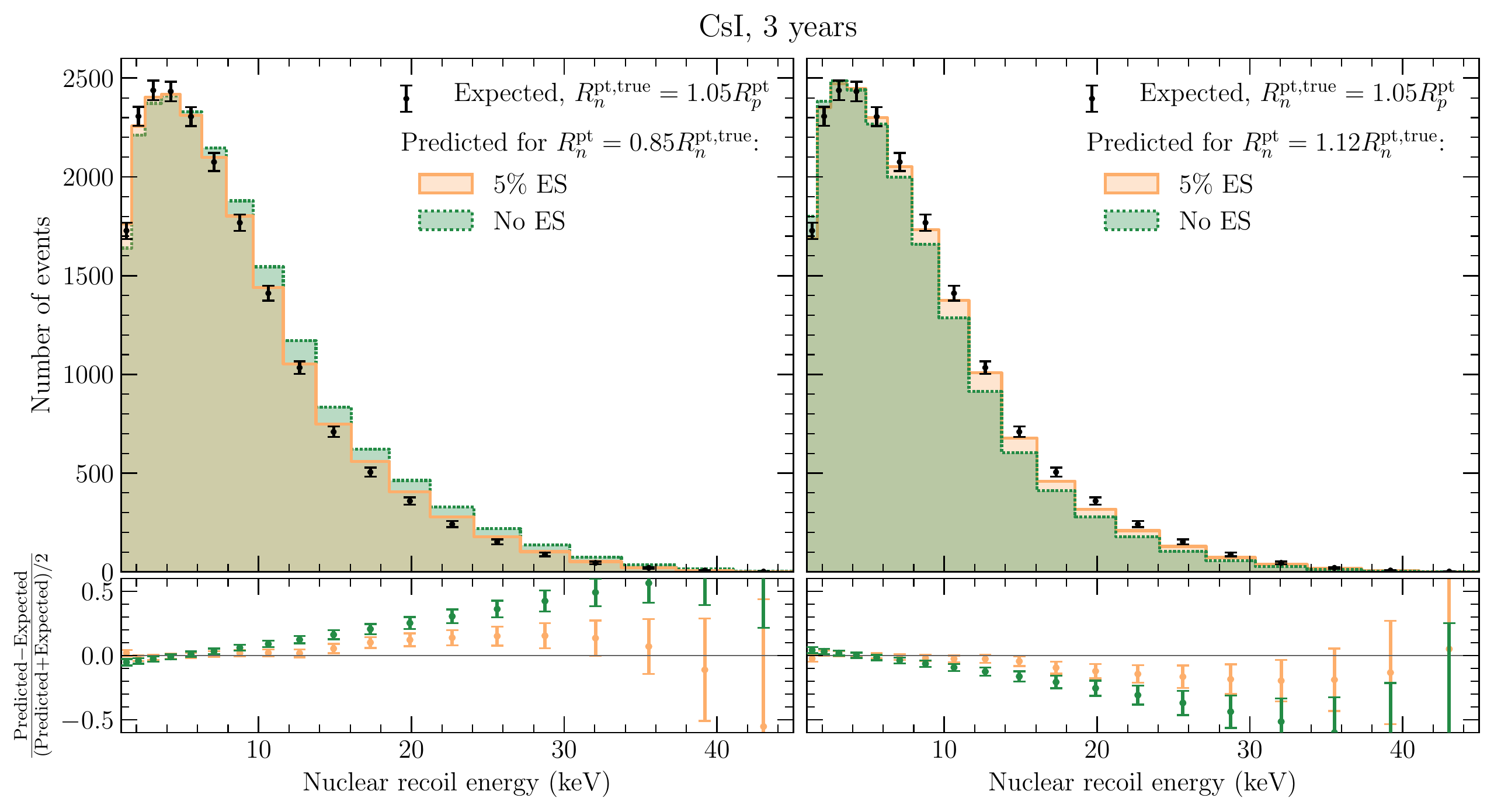}
\caption{Recoil energy spectrum of events in the cryogenic undoped CsI
  scintillator array.  The data points correspond to the simulated
  {\sl data} event rates for $R_n^\mathrm{pt,true}=1.05\, R_p$ and the
  error bars represent their statistical error. The colored histograms
  show the expected distributions for $R^{\rm pt}_n=0.85\, R^{\rm
    pt}_p$ (left panel) and $R^{\rm pt}_n=1.12\, R^{\rm pt}_p$ (right
  panel panel), with pulls chosen to optimize the fit to the simulated
  data. The different histograms correspond to different assumptions
  of the energy scale systematic uncertainty (see text for details).
  The lower panels show the relative difference in the event rates per
  bin, with error bars showing the corresponding
  uncertainties. \label{fig:spec-ESS} }
\end{figure}

Figure~\ref{fig:spec-ESS} illustrates the impact on the sensitivity due to the ES uncertainty. 
In both panels, the black dots show the simulated \emph{data} for the
CsI detector, for our assumed value of $R_n^\mathrm{pt,true}$. The
colored histograms, on the other hand, show the predicted spectra for
different test values of $R^{\rm pt}_n$ ($R^{\rm pt}_n < R_n^{\rm pt,true}$
in the left panel, while $R^{\rm pt}_n > R_n^{\rm pt,true}$ in the right panel), 
and for different assumptions
on the ES systematic error, after setting the nuisance parameters at the values which
give the best possible fit to the simulated data points.
The dotted green histograms have been obtained without an
ES uncertainty, and therefore show directly the impact of $R^{\rm pt}_n$ on the
event distributions: as expected from the analytic expressions in
Sec.~\ref{sec:framework}, an excess of events is observed for $R^{\rm
  pt}_n\leq R_n^\mathrm{pt,true}$ (left panel), particularly
relevant at high recoils, while the right panel shows the opposite
behaviour. Once an ES uncertainty is added, the fit will
try to vary the nuisance parameters in the $\chi^2$ to find a better
fit to the \emph{data}. As can be seen, the inclusion of this pull
term allows to induce apparent changes in the observed spectrum, which
mimic the impact of a different value of $R^{\rm pt}_n$ in the fit.
This is better appreciated in the lower panels, which show the
relative difference in the event rates per bin (with error bars
showing the corresponding uncertainties). As shown in the lower panels,
the inclusion of an energy-independent ES uncertainty significantly
relaxes the tension in the fit in both cases.

\begin{figure}[t]
\centering \includegraphics[width=\textwidth]{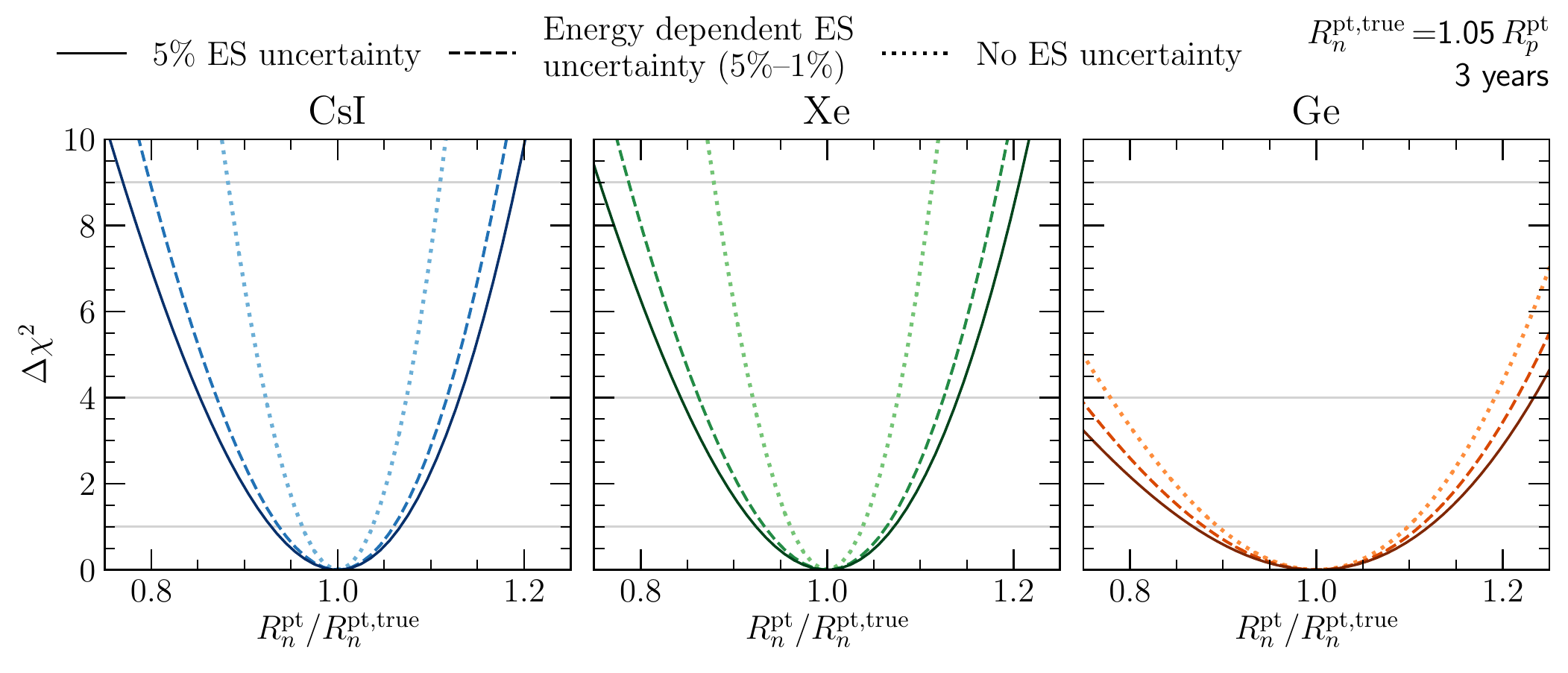}
\caption{$\Delta\chi^2$ as a function of the rms radius of the neutron
  distribution $R^{\rm pt}_n$ for three detectors proposed for a CE$\nu$NS
  experiment at the ESS: a cryogenic undoped CsI scintillator array
  (left panel) a high-pressure gaseous xenon chamber (central panel),
  and a low-threshold, multi-kg p-type point contact germanium
  detector (right panel).  In all cases an exposure of 3 years is
  assumed, together with a 10\% signal normalization uncertainty and a
  5\% background normalization uncertainty.  The three curves in each
  panel show our results for different assumptions for the systematic
  uncertainty in the energy scale reconstruction: negligible (dotted
  lines), energy-dependent, varying from 5\% at $T_{\mathrm{rec, th}}$ 
  to 1\% at $T_{\mathrm{rec, max}}$ (dashed lines), and 5\%
  energy-independent (solid lines).}
\label{fig:Rn-ESS}
\end{figure}

Finally Fig.~\ref{fig:Rn-ESS} shows the resulting
$\chi^2(R^{\rm pt}_n)$ for our ESS
simulations, as a function of $R^{\rm pt}_n/R_n^\mathrm{pt,true}$  for
the three different detectors as well as for the different assumptions of the
systematic uncertainties considered.
By construction, the $\chi^2(R^{\rm pt}_n)$ has its minimum at zero for
$R^{\rm pt}_n/R_n^\mathrm{pt,true} =1$ for all the curves, as expected. From the
comparison between the dotted lines in each panel and the dashed/solid
lines we immediately observe that the inclusion of an ES
uncertainty, even if smaller than the overall normalization
uncertainty, significantly spoils the sensitivity. This is expected,
since the main effect of $R^{\rm pt}_n$ is the distortion of the tail of the
recoil energy spectrum as described in Sec.~\ref{sec:framework} (see
also Fig.~\ref{fig:spec-ESS}), an effect that can be mimicked by an
ES uncertainty.  

From the comparison between the different panels in
Fig.~\ref{fig:Rn-ESS} we can also see how the sensitivity is
substantially worsened for lighter nuclei. This is expected \emph{a
  priori} since the signal statistics for \cevns grows quadratically
with the number of neutrons in the target nucleus, see Eqs.~\eqref{eq:xsec-short} and
\eqref{eq:RW-long}. Moreover, note that the values of $R^{\rm pt}_{p,n}$ (or,
equivalently, $R_W$) characterize the nuclear size, and
therefore are significantly smaller for Ge than for Xe or CsI. Thus, from
Eq.~\eqref{eq:FW-RW} one can see that for smaller nuclei the
dependence of the form factor with $Q^2$ will be very mild in these cases. Since the
sensitivity to $R^{\rm pt}_n$ comes precisely from the observation of a change
in the shape of the energy distribution, this automatically translates
into a worse sensitivity for smaller nuclei. In other words: the
sensitivity to $R^{\rm pt}_n$ comes from the fact that the wavelength of the
neutrino is of the order of the size of the nucleus, so it is
sensitive to the nuclear structure and, in particular, to the size of
the neutron distribution. However, as the size of the nucleus
decreases larger neutrino energies are required to probe the size of the nucleon distribution. 
This poses a challenge for CE$\nu$NS, for which
low momentum-transfers are required to maintain coherence.

\section{Summary and Conclusions} 
\label{sec:conclusions}
Coherent elastic neutrino-nucleus scattering (\cevns) probes
the weak form factor of the nucleus and provides us with
information of its weak charge distribution.  In particular, from
\cevns data it is possible to determine the rms radius of
the neutron distribution of the target nuclei.
The difference with the rms proton distribution---the neutron skin thickness---is
crucial in a broad spectrum of nuclear physics and astrophysics, including nuclear structure, nuclear matter, and neutron stars.

First, as reference, and for comparison with the existing
literature we have  performed an analysis of the present results
from COHERENT experiment using the detailed timing and energy
information of their data for CsI.
We found that COHERENT provides a 1$\sigma$ uncertainty
on the rms radius of the neutron distribution which ranges from 11\%
to 16\%, with an additional 11\% uncertainty that depends on the
choice of QF and SS background modeling.  In particular, we get the
following determination of $R^{\rm pt}_n$ (at $1\sigma$) from the different
analyses:
\begin{eqnarray}
  R^{\rm pt}_n=6.28^{+0.77}_{-0.85}\;{\rm fm} & &{\rm For\; Data\;Release\;t+E} \\
  R^{\rm pt}_n=5.80^{+0.89}_{-0.93}\; {\rm fm}& &{\rm For\;
    Data\;Release\;t+E\; +\; new\; Background}\\
  R^{\rm pt}_n=5.96^{+0.57}_{-0.59} \;{\rm fm}& &{\rm For\; Our\;Fit\;t+E\;D} \\
  R^{\rm pt}_n  =4.99^{+0.65}_{-0.73} \;{\rm fm}& &{\rm For\; Our\;Fit\; t+E\;C}
\end{eqnarray}

\begin{figure}[t]
	\centering \includegraphics [width=\textwidth]{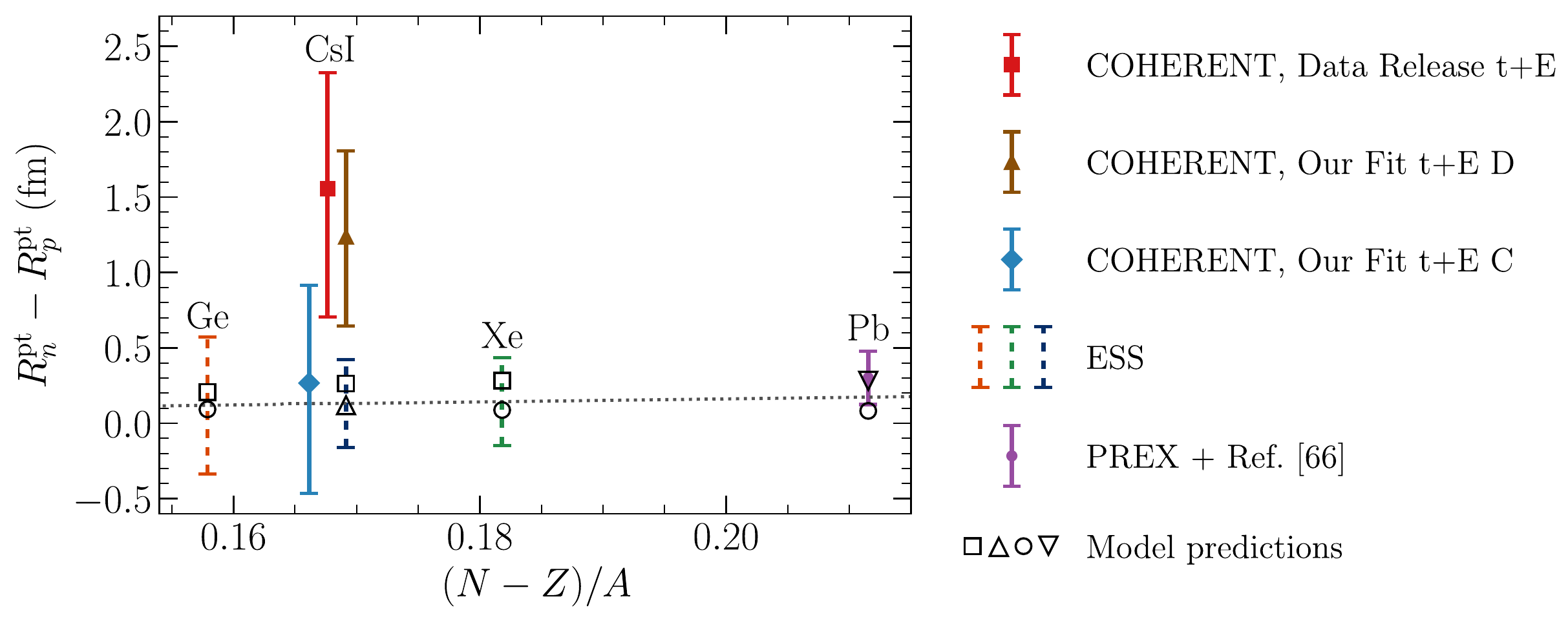}
	\caption{Compilation of our results for the present determination and
		future sensitivity  of the neutron skin thickness from \cevns experiments
		for the isospin asymmetric nucleus considered in this work.
		For clarity, the red and blue points corresponding to two analyses of
		COHERENT have been slightly
		displaced horizontally.
		For comparison we show the result for $^{208}$Pb derived from the PREX
		measurement with the analysis in Ref.~\cite{Horowitz:2012tj}, and 
		the best-fit to the indirect determination from antiprotonic atom x-ray data for a variety of nuclei 
		(dashed line), taken from Ref.~\cite{Trzcinska:2001sy}. The hollow markers represent the range of 
		predictions in a variety of models, see text for details. For concreteness, the expected results from the ESS 
		have been centered on the dashed line, and correspond to our analysis with an energy-dependent ES 
		uncertainty ranging from 1\% to 5\%. }
	\label{fig:rnglobal}
\end{figure}

This is to be compared with the corresponding values for the proton
distributions  for Cs, $R_p^{\rm pt,Cs} = 4.75$~fm, and I, $R_p^{\rm pt,I} = 4.70$~fm.
In Fig.~\ref{fig:rnglobal}
we plot the corresponding results as
the inferred neutron skin thickness
by subtracting the average
$R_p^{\rm pt,CsI} = 4.725$~fm.
Despite the relatively poor precision, it is important to stress that
these are the only direct probes available for this observable. Even
within present uncertainties, it is interesting to point
out that the results from the data release (original background) and
our fit with the Duke QF lead to values of the neutron skin that
exceed those predicted by theoretical nuclear models~\cite{Cadeddu:2017etk, HoMeSh:prep}. In
contrast, the analysis with the Chicago QF yields a thinner neutron skin,
consistent with calculations. However, note that the Chicago QF 1$\sigma$ errors reach negative values for the 
inferred neutron skin thickness,
an unexpected result for a neutron-rich nucleus~\cite{BohrMottelson} not predicted by any nuclear model~\cite{Centelles:2008vu,Erler2012,Hagen:2015yea,Cadeddu:2017etk,Co:2020gwl}.

For comparison, Fig.~\ref{fig:rnglobal} shows the neutron skin thickness for $^{208}$Pb, derived from the
analysis in Ref.~\cite{Horowitz:2012tj} of the results of the Lead Radius Experiment (PREX) experiment~\cite{Abrahamyan:2012gp} on parity-violation in electron scattering. Let us stress  that although parity-violation is a weak measurement
(and therefore it does probe the neutron distribution of the nucleus),
PREX does so only at a fixed momentum transfer.
The extraction of the point-neutron radius and the corresponding skin thickness
from the asymmetry cross-section measured by PREX contains certain model dependence,
as can be observed by the slightly different results presented by Refs.~\cite{Abrahamyan:2012gp} and~\cite{Horowitz:2012tj}.
In this figure we also show the best-fit for the neutron skin obtained from 
its indirect determination from antiprotonic X-ray data for a variety of nuclei (dashed line), taken from Ref.~\cite{Trzcinska:2001sy}. 
Finally, the hollow markers show 
the predictions for a variety of nuclear models as extracted from
Refs.~\cite{Co:2020gwl, Hoferichter:2018acd, Cadeddu:2017etk, HoMeSh:prep}.

Regarding the future sensitivity at a \cevns experiment using the ESS
as the neutrino source we have focused on the proposed detectors using
three different detector materials: CsI, Xe and Ge. With the large
statistics expected, variations of $R^{\rm pt}_n$ of ${\cal O}$(few \%) can lead
to observable distortions in the recoil energy spectrum provided it
can be measured with enough precision. We have explored this effect
and obtained the following expected precision for the determination  of 
$R^{\rm pt}_n$ at 1$\sigma$, for the different detectors and three
choices of energy scale (ES) systematic uncertainties, namely, negligible ES
uncertainty (5\% to 1\% ES uncertainty)
[5\% ES uncertainty]:
\begin{eqnarray}
\label{eq:finalRn}
  {\rm CsI:} & &\sigma(R^{\rm pt}_n)/ R^{\rm pt,true}_n = 4\% (6\%) [7\%] \nonumber \\
  {\rm Xe:} && \sigma(R^{\rm pt}_n)/ R^{\rm pt,true}_n = 4\% (6\%) [7\%]\\ 
  {\rm Ge:} && \sigma(R^{\rm pt}_n)/ R^{\rm pt,true}_n = 10\% (11\%) [12\%] \nonumber
\end{eqnarray}
Our results for the expected precision at the ESS are also shown in Fig.~\ref{fig:rnglobal}, for the
analysis with (5\% to 1\% ES uncertainty).  

While the uncertainties in Eq.~\eqref{eq:finalRn} lead to a range of
values of the neutron skin thickness which exceeds the typical spread of nuclear
structure results using different models,
$\sim (0.1-0.3)\,$fm~\cite{Centelles:2008vu,Cadeddu:2017etk,Co:2020gwl},
Fig.~\ref{fig:rnglobal} shows that \cevns can provide useful
constraints on the neutron skin of heavy, neutron-rich nuclei.  In fact
\cevns gives the most direct measurement of the neutron
skin, with nuclear-model independent error bars, bringing
the opportunity to measure it for different nuclei relatively easily.
Combined with future improvements beyond the ESS, these advantages place
\cevns in a position towards determining unambiguously the size of the
neutron distribution with model-independent uncertainties.

\section*{Acknowledgments}

We thank J.~I.~Collar, M.~Hoferichter, F.~Monrabal and A.~Schwenk for
useful discussions.  This work was supported by the MINECO grants
FPA2016-76005-C2-1-P and FIS2017-87534-P, by the MINECO FEDER/UE
grants FPA2015-65929-P, FPA2016-78645-P and FPA2017-85985-P, by
PROMETEO/2019/083, by USA-NSF grants PHY-1620628, by AGAUR
(Generalitat de Catalunya) grant 2017-SGR-929, and by the European ITN
project HiDDeN (H2020-MSCA-ITN-2019//860881-HIDDeN).  IE acknowledges
support from the FPU program fellowship FPU15/0369. PC and JM
acknowledge support from the Spanish MICINN through the ``Ram\'on y
Cajal'' program with grants RYC2018-024240-I (PC) and RYC-2017-22781
(JM).  The authors acknowledge the support of the Spanish Agencia
Estatal de Investigacion through the grant ``IFT Centro de Excelencia
Severo Ochoa SEV-2016-0597''.

\bibliographystyle{JHEP}
\bibliography{Bibliography}

\providecommand{\href}[2]{#2}\begingroup\raggedright\begin{thebibliography}{10}

\bibitem{freedman}
D.~Z. Freedman, {\it {Coherent Neutrino Nucleus Scattering as a Probe of the
  Weak Neutral Current}},  {\em Phys. Rev.} {\bf D9} (1974) 1389--1392.

\bibitem{Akimov:2017ade}
{\bf COHERENT} Collaboration, D.~Akimov {\em et~al.}, {\it {Observation of
  Coherent Elastic Neutrino-Nucleus Scattering}},  {\em Science} {\bf 357}
  (2017), no.~6356 1123--1126, [\href{http://xxx.lanl.gov/abs/1708.01294}{{\tt
  1708.01294}}].

\bibitem{Akimov:2018vzs}
{\bf COHERENT} Collaboration, D.~Akimov {\em et~al.}, {\it {COHERENT
  Collaboration data release from the first observation of coherent elastic
  neutrino-nucleus scattering}},
  \href{http://xxx.lanl.gov/abs/1804.09459}{{\tt 1804.09459}}.

\bibitem{Akimov:2020pdx}
{\bf COHERENT} Collaboration, D.~Akimov {\em et~al.}, {\it {First Detection of
  Coherent Elastic Neutrino-Nucleus Scattering on Argon}},
  \href{http://xxx.lanl.gov/abs/2003.10630}{{\tt 2003.10630}}.

\bibitem{nsi2}
P.~Coloma, M.~C. Gonzalez-Garcia, M.~Maltoni, and T.~Schwetz, {\it {COHERENT
  Enlightenment of the Neutrino Dark Side}},  {\em Phys. Rev.} {\bf D96}
  (2017), no.~11 115007, [\href{http://xxx.lanl.gov/abs/1708.02899}{{\tt
  1708.02899}}].

\bibitem{nsi1}
J.~B. Dent, B.~Dutta, S.~Liao, J.~L. Newstead, L.~E. Strigari, and J.~W.
  Walker, {\it {Probing light mediators at ultralow threshold energies with
  coherent elastic neutrino-nucleus scattering}},  {\em Phys. Rev.} {\bf D96}
  (2017), no.~9 095007, [\href{http://xxx.lanl.gov/abs/1612.06350}{{\tt
  1612.06350}}].

\bibitem{nsi3}
J.~Liao and D.~Marfatia, {\it {COHERENT constraints on nonstandard neutrino
  interactions}},  {\em Phys. Lett.} {\bf B775} (2017) 54--57,
  [\href{http://xxx.lanl.gov/abs/1708.04255}{{\tt 1708.04255}}].

\bibitem{nsi4}
J.~B. Dent, B.~Dutta, S.~Liao, J.~L. Newstead, L.~E. Strigari, and J.~W.
  Walker, {\it {Accelerator and reactor complementarity in coherent
  neutrino-nucleus scattering}},  {\em Phys. Rev.} {\bf D97} (2018), no.~3
  035009, [\href{http://xxx.lanl.gov/abs/1711.03521}{{\tt 1711.03521}}].

\bibitem{nsi5}
Y.~Farzan, M.~Lindner, W.~Rodejohann, and X.-J. Xu, {\it {Probing neutrino
  coupling to a light scalar with coherent neutrino scattering}},  {\em JHEP}
  {\bf 05} (2018) 066, [\href{http://xxx.lanl.gov/abs/1802.05171}{{\tt
  1802.05171}}].

\bibitem{nsi6}
M.~Abdullah, J.~B. Dent, B.~Dutta, G.~L. Kane, S.~Liao, and L.~E. Strigari,
  {\it {Coherent elastic neutrino nucleus scattering as a probe of a Z through
  kinetic and mass mixing effects}},  {\em Phys. Rev.} {\bf D98} (2018), no.~1
  015005, [\href{http://xxx.lanl.gov/abs/1803.01224}{{\tt 1803.01224}}].

\bibitem{nsi7}
I.~Esteban, M.~C. Gonzalez-Garcia, M.~Maltoni, I.~Martinez-Soler, and
  J.~Salvado, {\it {Updated Constraints on Non-Standard Interactions from
  Global Analysis of Oscillation Data}},  {\em JHEP} {\bf 08} (2018) 180,
  [\href{http://xxx.lanl.gov/abs/1805.04530}{{\tt 1805.04530}}].

\bibitem{nsi8}
D.~Aristizabal~Sierra, V.~De~Romeri, and N.~Rojas, {\it {COHERENT analysis of
  neutrino generalized interactions}},  {\em Phys. Rev.} {\bf D98} (2018)
  075018, [\href{http://xxx.lanl.gov/abs/1806.07424}{{\tt 1806.07424}}].

\bibitem{nsi9}
I.~M. Shoemaker, {\it {COHERENT search strategy for beyond standard model
  neutrino interactions}},  {\em Phys. Rev.} {\bf D95} (2017), no.~11 115028,
  [\href{http://xxx.lanl.gov/abs/1703.05774}{{\tt 1703.05774}}].

\bibitem{nsi10}
P.~Coloma, P.~B. Denton, M.~C. Gonzalez-Garcia, M.~Maltoni, and T.~Schwetz,
  {\it {Curtailing the Dark Side in Non-Standard Neutrino Interactions}},  {\em
  JHEP} {\bf 04} (2017) 116, [\href{http://xxx.lanl.gov/abs/1701.04828}{{\tt
  1701.04828}}].

\bibitem{Giunti:2019xpr}
C.~Giunti, {\it {General COHERENT constraints on neutrino nonstandard
  interactions}},  {\em Phys. Rev. D} {\bf 101} (2020), no.~3 035039,
  [\href{http://xxx.lanl.gov/abs/1909.00466}{{\tt 1909.00466}}].

\bibitem{Denton:2018xmq}
P.~B. Denton, Y.~Farzan, and I.~M. Shoemaker, {\it {Testing large non-standard
  neutrino interactions with arbitrary mediator mass after COHERENT data}},
  {\em JHEP} {\bf 07} (2018) 037,
  [\href{http://xxx.lanl.gov/abs/1804.03660}{{\tt 1804.03660}}].

\bibitem{Coloma:2019mbs}
P.~Coloma, I.~Esteban, M.~Gonzalez-Garcia, and M.~Maltoni, {\it {Improved
  global fit to Non-Standard neutrino Interactions using COHERENT energy and
  timing data}},  {\em JHEP} {\bf 02} (2020) 023,
  [\href{http://xxx.lanl.gov/abs/1911.09109}{{\tt 1911.09109}}].

\bibitem{Flores:2020lji}
L.~Flores, N.~Nath, and E.~Peinado, {\it {Non-standard neutrino interactions in
  $U(1)'$ model after COHERENT data}},  {\em JHEP} {\bf 06} (2020) 045,
  [\href{http://xxx.lanl.gov/abs/2002.12342}{{\tt 2002.12342}}].

\bibitem{em1}
D.~K. Papoulias and T.~S. Kosmas, {\it {COHERENT constraints to conventional
  and exotic neutrino physics}},  {\em Phys. Rev.} {\bf D97} (2018), no.~3
  033003, [\href{http://xxx.lanl.gov/abs/1711.09773}{{\tt 1711.09773}}].

\bibitem{em2}
J.~Billard, J.~Johnston, and B.~J. Kavanagh, {\it {Prospects for exploring New
  Physics in Coherent Elastic Neutrino-Nucleus Scattering}},  {\em JCAP} {\bf
  1811} (2018), no.~11 016, [\href{http://xxx.lanl.gov/abs/1805.01798}{{\tt
  1805.01798}}].

\bibitem{em3}
M.~Cadeddu, C.~Giunti, K.~A. Kouzakov, Y.~F. Li, A.~I. Studenikin, and Y.~Y.
  Zhang, {\it {Neutrino Charge Radii from COHERENT Elastic Neutrino-Nucleus
  Scattering}},  {\em Phys. Rev.} {\bf D98} (2018), no.~11 113010,
  [\href{http://xxx.lanl.gov/abs/1810.05606}{{\tt 1810.05606}}].

\bibitem{em4}
O.~G. Miranda, D.~K. Papoulias, M.~Tórtola, and J.~W.~F. Valle, {\it {Probing
  neutrino transition magnetic moments with coherent elastic neutrino-nucleus
  scattering}},  {\em JHEP} {\bf 07} (2019) 103,
  [\href{http://xxx.lanl.gov/abs/1905.03750}{{\tt 1905.03750}}].

\bibitem{Papoulias:2019txv}
D.~K. Papoulias, {\it {COHERENT constraints after the Chicago-3 quenching
  factor measurement}},  \href{http://xxx.lanl.gov/abs/1907.11644}{{\tt
  1907.11644}}.

\bibitem{ste1}
T.~S. Kosmas, D.~K. Papoulias, M.~Tortola, and J.~W.~F. Valle, {\it {Probing
  light sterile neutrino signatures at reactor and Spallation Neutron Source
  neutrino experiments}},  {\em Phys. Rev.} {\bf D96} (2017), no.~6 063013,
  [\href{http://xxx.lanl.gov/abs/1703.00054}{{\tt 1703.00054}}].

\bibitem{carlos}
C.~Blanco, D.~Hooper, and P.~Machado, {\it {Constraining Sterile Neutrino
  Interpretations of the LSND and MiniBooNE Anomalies with Coherent Neutrino
  Scattering Experiments}},  \href{http://xxx.lanl.gov/abs/1901.08094}{{\tt
  1901.08094}}.

\bibitem{dm1}
S.-F. Ge and I.~M. Shoemaker, {\it {Constraining Photon Portal Dark Matter with
  Texono and Coherent Data}},  {\em JHEP} {\bf 11} (2018) 066,
  [\href{http://xxx.lanl.gov/abs/1710.10889}{{\tt 1710.10889}}].

\bibitem{dm2}
V.~Brdar, W.~Rodejohann, and X.-J. Xu, {\it {Producing a new Fermion in
  Coherent Elastic Neutrino-Nucleus Scattering: from Neutrino Mass to Dark
  Matter}},  {\em JHEP} {\bf 12} (2018) 024,
  [\href{http://xxx.lanl.gov/abs/1810.03626}{{\tt 1810.03626}}].

\bibitem{dm3}
B.~Dutta, D.~Kim, S.~Liao, J.-C. Park, S.~Shin, and L.~E. Strigari, {\it {Dark
  matter signals from timing spectra at neutrino experiments}},
  \href{http://xxx.lanl.gov/abs/1906.10745}{{\tt 1906.10745}}.

\bibitem{wma1}
B.~C. Cañas, E.~A. Garcés, O.~G. Miranda, and A.~Parada, {\it {Future
  perspectives for a weak mixing angle measurement in coherent elastic neutrino
  nucleus scattering experiments}},  {\em Phys. Lett.} {\bf B784} (2018)
  159--162, [\href{http://xxx.lanl.gov/abs/1806.01310}{{\tt 1806.01310}}].

\bibitem{wma2}
M.~Cadeddu and F.~Dordei, {\it {Reinterpreting the weak mixing angle from
  atomic parity violation in view of the Cs neutron rms radius measurement from
  COHERENT}},  {\em Phys. Rev.} {\bf D99} (2019), no.~3 033010,
  [\href{http://xxx.lanl.gov/abs/1808.10202}{{\tt 1808.10202}}].

\bibitem{wma3}
X.-R. Huang and L.-W. Chen, {\it {Neutron Skin in CsI and Low-Energy Effective
  Weak Mixing Angle from COHERENT Data}},  {\em Phys. Rev.} {\bf D100} (2019),
  no.~7 071301, [\href{http://xxx.lanl.gov/abs/1902.07625}{{\tt 1902.07625}}].

\bibitem{Cadeddu:2017etk}
M.~Cadeddu, C.~Giunti, Y.~Li, and Y.~Zhang, {\it {Average CsI neutron density
  distribution from COHERENT data}},  {\em Phys. Rev. Lett.} {\bf 120} (2018),
  no.~7 072501, [\href{http://xxx.lanl.gov/abs/1710.02730}{{\tt 1710.02730}}].

\bibitem{Ciuffoli:2018qem}
E.~Ciuffoli, J.~Evslin, Q.~Fu, and J.~Tang, {\it {Extracting nuclear form
  factors with coherent neutrino scattering}},  {\em Phys. Rev. D} {\bf 97}
  (2018), no.~11 113003, [\href{http://xxx.lanl.gov/abs/1801.02166}{{\tt
  1801.02166}}].

\bibitem{Cadeddu:2018izq}
M.~Cadeddu and F.~Dordei, {\it {Reinterpreting the weak mixing angle from
  atomic parity violation in view of the Cs neutron rms radius measurement from
  COHERENT}},  {\em Phys. Rev. D} {\bf 99} (2019), no.~3 033010,
  [\href{http://xxx.lanl.gov/abs/1808.10202}{{\tt 1808.10202}}].

\bibitem{Cadeddu:2019eta}
M.~Cadeddu, F.~Dordei, C.~Giunti, Y.~Li, and Y.~Zhang, {\it {Neutrino,
  electroweak, and nuclear physics from COHERENT elastic neutrino-nucleus
  scattering with refined quenching factor}},  {\em Phys. Rev. D} {\bf 101}
  (2020), no.~3 033004, [\href{http://xxx.lanl.gov/abs/1908.06045}{{\tt
  1908.06045}}].

\bibitem{Papoulias:2019lfi}
D.~Papoulias, T.~Kosmas, R.~Sahu, V.~Kota, and M.~Hota, {\it {Constraining
  nuclear physics parameters with current and future COHERENT data}},  {\em
  Phys. Lett. B} {\bf 800} (2020) 135133,
  [\href{http://xxx.lanl.gov/abs/1903.03722}{{\tt 1903.03722}}].

\bibitem{Khan:2019cvi}
A.~N. Khan and W.~Rodejohann, {\it {New physics from COHERENT data with an
  improved quenching factor}},  {\em Phys. Rev. D} {\bf 100} (2019), no.~11
  113003, [\href{http://xxx.lanl.gov/abs/1907.12444}{{\tt 1907.12444}}].

\bibitem{Huang:2019ene}
X.-R. Huang and L.-W. Chen, {\it {Neutron Skin in CsI and Low-Energy Effective
  Weak Mixing Angle from COHERENT Data}},  {\em Phys. Rev. D} {\bf 100} (2019),
  no.~7 071301, [\href{http://xxx.lanl.gov/abs/1902.07625}{{\tt 1902.07625}}].

\bibitem{Canas:2019fjw}
B.~Canas, E.~Garces, O.~Miranda, A.~Parada, and G.~Sanchez~Garcia, {\it
  {Interplay between nonstandard and nuclear constraints in coherent elastic
  neutrino-nucleus scattering experiments}},  {\em Phys. Rev. D} {\bf 101}
  (2020), no.~3 035012, [\href{http://xxx.lanl.gov/abs/1911.09831}{{\tt
  1911.09831}}].

\bibitem{Cadeddu:2020lky}
M.~Cadeddu, F.~Dordei, C.~Giunti, Y.~Li, E.~Picciau, and Y.~Zhang, {\it
  {Physics results from the first COHERENT observation of CE$\nu$NS in argon
  and their combination with cesium-iodide data}},
  \href{http://xxx.lanl.gov/abs/2005.01645}{{\tt 2005.01645}}.

\bibitem{Miranda:2020tif}
O.~Miranda, D.~Papoulias, G.~S. Garcia, O.~Sanders, M.~Tórtola, and J.~Valle,
  {\it {Implications of the first detection of coherent elastic
  neutrino-nucleus scattering (CEvNS) with Liquid Argon}},
  \href{http://xxx.lanl.gov/abs/2003.12050}{{\tt 2003.12050}}.

\bibitem{Donnelly:1984rg}
T.~Donnelly and I.~Sick, {\it {ELASTIC MAGNETIC ELECTRON SCATTERING FROM
  NUCLEI}},  {\em Rev. Mod. Phys.} {\bf 56} (1984) 461--566.

\bibitem{Angeli:2013epw}
I.~Angeli and K.~Marinova, {\it {Table of experimental nuclear ground state
  charge radii: An update}},  {\em Atom. Data Nucl. Data Tabl.} {\bf 99}
  (2013), no.~1 69--95.

\bibitem{Abrahamyan:2012gp}
S.~Abrahamyan {\em et~al.}, {\it {Measurement of the Neutron Radius of 208Pb
  Through Parity-Violation in Electron Scattering}},  {\em Phys. Rev. Lett.}
  {\bf 108} (2012) 112502, [\href{http://xxx.lanl.gov/abs/1201.2568}{{\tt
  1201.2568}}].

\bibitem{GarciaRecio:1991wk}
C.~Garcia-Recio, J.~Nieves, and E.~Oset, {\it {Neutron distributions from
  pionic atoms}},  {\em Nucl. Phys. A} {\bf 547} (1992) 473--487.

\bibitem{Suzuki:1995yc}
T.~Suzuki {\em et~al.}, {\it {Neutron skin of Na isotopes studied via the
  interaction cross-sections}},  {\em Phys. Rev. Lett.} {\bf 75} (1995)
  3241--3244.

\bibitem{Clark:2002se}
B.~Clark, L.~Kerr, and S.~Hama, {\it {Neutron densities from a global analysis
  of medium-energy proton nucleus elastic scattering}},  {\em Phys. Rev. C}
  {\bf 67} (2003) 054605, [\href{http://xxx.lanl.gov/abs/nucl-th/0209052}{{\tt
  nucl-th/0209052}}].

\bibitem{Trzcinska:2001sy}
A.~Trzcinska, J.~Jastrzebski, P.~Lubinski, F.~Hartmann, R.~Schmidt, T.~von
  Egidy, and B.~Klos, {\it {Neutron density distributions deduced from
  anti-protonic atoms}},  {\em Phys. Rev. Lett.} {\bf 87} (2001) 082501.

\bibitem{Lapoux:2016exf}
V.~Lapoux, V.~Somà, C.~Barbieri, H.~Hergert, J.~Holt, and S.~Stroberg, {\it
  {Radii and Binding Energies in Oxygen Isotopes: A Challenge for Nuclear
  Forces}},  {\em Phys. Rev. Lett.} {\bf 117} (2016), no.~5 052501,
  [\href{http://xxx.lanl.gov/abs/1605.07885}{{\tt 1605.07885}}].

\bibitem{Tarbert:2013jze}
C.~Tarbert {\em et~al.}, {\it {Neutron skin of $^{208}$Pb from Coherent Pion
  Photoproduction}},  {\em Phys. Rev. Lett.} {\bf 112} (2014), no.~24 242502,
  [\href{http://xxx.lanl.gov/abs/1311.0168}{{\tt 1311.0168}}].

\bibitem{Horowitz:1999fk}
C.~Horowitz, S.~Pollock, P.~Souder, and R.~Michaels, {\it {Parity violating
  measurements of neutron densities}},  {\em Phys. Rev. C} {\bf 63} (2001)
  025501, [\href{http://xxx.lanl.gov/abs/nucl-th/9912038}{{\tt
  nucl-th/9912038}}].

\bibitem{Brown:2008ib}
B.~Brown, A.~Derevianko, and V.~Flambaum, {\it {Calculations of the neutron
  skin and its effect in atomic parity violation}},  {\em Phys. Rev. C} {\bf
  79} (2009) 035501, [\href{http://xxx.lanl.gov/abs/0804.4315}{{\tt
  0804.4315}}].

\bibitem{Dzuba:2012kx}
V.~Dzuba, J.~Berengut, V.~Flambaum, and B.~Roberts, {\it {Revisiting parity
  non-conservation in cesium}},  {\em Phys. Rev. Lett.} {\bf 109} (2012)
  203003, [\href{http://xxx.lanl.gov/abs/1207.5864}{{\tt 1207.5864}}].

\bibitem{Viatkina:2019wsz}
A.~Viatkina, D.~Antypas, M.~Kozlov, D.~Budker, and V.~Flambaum, {\it
  {Dependence of atomic parity-violation effects on neutron skins and new
  physics}},  {\em Phys. Rev. C} {\bf 100} (2019), no.~3 034318,
  [\href{http://xxx.lanl.gov/abs/1903.00123}{{\tt 1903.00123}}].

\bibitem{Erler2012}
J.~Erler, N.~Birge, M.~Kortelainen, W.~Nazarewicz, E.~Olsen, P.~A. M., and
  M.~Stoitsov, {\it {The limits of the nuclear landscape}},  {\em Nature} {\bf
  486} (2012) 509.

\bibitem{Tanihata:2013jwa}
I.~Tanihata, H.~Savajols, and R.~Kanungo, {\it {Recent experimental progress in
  nuclear halo structure studies}},  {\em Prog. Part. Nucl. Phys.} {\bf 68}
  (2013) 215--313.

\bibitem{Hagen:2015yea}
G.~Hagen {\em et~al.}, {\it {Neutron and weak-charge distributions of the
  $^{48}$Ca nucleus}},  {\em Nature Phys.} {\bf 12} (2015), no.~2 186--190,
  [\href{http://xxx.lanl.gov/abs/1509.07169}{{\tt 1509.07169}}].

\bibitem{Brown:2000pd}
B.~Brown, {\it {Neutron radii in nuclei and the neutron equation of state}},
  {\em Phys. Rev. Lett.} {\bf 85} (2000) 5296--5299.

\bibitem{Horowitz:2000xj}
C.~Horowitz and J.~Piekarewicz, {\it {Neutron star structure and the neutron
  radius of Pb-208}},  {\em Phys. Rev. Lett.} {\bf 86} (2001) 5647,
  [\href{http://xxx.lanl.gov/abs/astro-ph/0010227}{{\tt astro-ph/0010227}}].

\bibitem{Centelles:2008vu}
M.~Centelles, X.~Roca-Maza, X.~Vinas, and M.~Warda, {\it {Nuclear symmetry
  energy probed by neutron skin thickness of nuclei}},  {\em Phys. Rev. Lett.}
  {\bf 102} (2009) 122502, [\href{http://xxx.lanl.gov/abs/0806.2886}{{\tt
  0806.2886}}].

\bibitem{Tsang:2012se}
M.~Tsang {\em et~al.}, {\it {Constraints on the symmetry energy and neutron
  skins from experiments and theory}},  {\em Phys. Rev. C} {\bf 86} (2012)
  015803, [\href{http://xxx.lanl.gov/abs/1204.0466}{{\tt 1204.0466}}].

\bibitem{Lattimer:2004pg}
J.~Lattimer and M.~Prakash, {\it {The physics of neutron stars}},  {\em
  Science} {\bf 304} (2004) 536--542,
  [\href{http://xxx.lanl.gov/abs/astro-ph/0405262}{{\tt astro-ph/0405262}}].

\bibitem{Lattimer:2006xb}
J.~M. Lattimer and M.~Prakash, {\it {Neutron Star Observations: Prognosis for
  Equation of State Constraints}},  {\em Phys. Rept.} {\bf 442} (2007)
  109--165, [\href{http://xxx.lanl.gov/abs/astro-ph/0612440}{{\tt
  astro-ph/0612440}}].

\bibitem{Baxter:2019mcx}
D.~Baxter {\em et~al.}, {\it {Coherent Elastic Neutrino-Nucleus Scattering at
  the European Spallation Source}},  {\em JHEP} {\bf 02} (2020) 123,
  [\href{http://xxx.lanl.gov/abs/1911.00762}{{\tt 1911.00762}}].

\bibitem{Barranco:2005yy}
J.~Barranco, O.~Miranda, and T.~Rashba, {\it {Probing new physics with coherent
  neutrino scattering off nuclei}},  {\em JHEP} {\bf 12} (2005) 021,
  [\href{http://xxx.lanl.gov/abs/hep-ph/0508299}{{\tt hep-ph/0508299}}].

\bibitem{Klos:2013rwa}
P.~Klos, J.~Menéndez, D.~Gazit, and A.~Schwenk, {\it {Large-scale nuclear
  structure calculations for spin-dependent WIMP scattering with chiral
  effective field theory currents}},  {\em Phys. Rev. D} {\bf 88} (2013), no.~8
  083516, [\href{http://xxx.lanl.gov/abs/1304.7684}{{\tt 1304.7684}}].
  [Erratum: Phys.Rev.D 89, 029901 (2014)].

\bibitem{Tanabashi:2018oca}
{\bf Particle Data Group} Collaboration, M.~Tanabashi {\em et~al.}, {\it
  {Review of Particle Physics}},  {\em Phys. Rev. D} {\bf 98} (2018), no.~3
  030001. 2019 update.

\bibitem{Erler:2013xha}
J.~Erler and S.~Su, {\it {The Weak Neutral Current}},  {\em Prog. Part. Nucl.
  Phys.} {\bf 71} (2013) 119--149,
  [\href{http://xxx.lanl.gov/abs/1303.5522}{{\tt 1303.5522}}].

\bibitem{Friar:1997js}
J.~L. Friar, J.~Martorell, and D.~Sprung, {\it {Nuclear sizes and the isotope
  shift}},  {\em Phys. Rev. A} {\bf 56} (1997) 4579--4586,
  [\href{http://xxx.lanl.gov/abs/nucl-th/9707016}{{\tt nucl-th/9707016}}].

\bibitem{Horowitz:2012we}
C.~Horowitz and J.~Piekarewicz, {\it {Impact of spin-orbit currents on the
  electroweak skin of neutron-rich nuclei}},  {\em Phys. Rev. C} {\bf 86}
  (2012) 045503, [\href{http://xxx.lanl.gov/abs/1208.2249}{{\tt 1208.2249}}].

\bibitem{Horowitz:2012tj}
C.~Horowitz {\em et~al.}, {\it {Weak charge form factor and radius of 208Pb
  through parity violation in electron scattering}},  {\em Phys. Rev. C} {\bf
  85} (2012) 032501, [\href{http://xxx.lanl.gov/abs/1202.1468}{{\tt
  1202.1468}}].

\bibitem{Hoferichter:2016nvd}
M.~Hoferichter, P.~Klos, J.~Menéndez, and A.~Schwenk, {\it {Analysis
  strategies for general spin-independent WIMP-nucleus scattering}},  {\em
  Phys. Rev.} {\bf D94} (2016), no.~6 063505,
  [\href{http://xxx.lanl.gov/abs/1605.08043}{{\tt 1605.08043}}].

\bibitem{Hoferichter:2018acd}
M.~Hoferichter, P.~Klos, J.~Menéndez, and A.~Schwenk, {\it {Nuclear structure
  factors for general spin-independent WIMP-nucleus scattering}},  {\em Phys.
  Rev.} {\bf D99} (2019), no.~5 055031,
  [\href{http://xxx.lanl.gov/abs/1812.05617}{{\tt 1812.05617}}].

\bibitem{Helm:1956zz}
R.~H. Helm, {\it {Inelastic and Elastic Scattering of 187-Mev Electrons from
  Selected Even-Even Nuclei}},  {\em Phys. Rev.} {\bf 104} (1956) 1466--1475.

\bibitem{Piekarewicz:2016vbn}
J.~Piekarewicz, A.~Linero, P.~Giuliani, and E.~Chicken, {\it {Power of two:
  Assessing the impact of a second measurement of the weak-charge form factor
  of $^{208}$Pb}},  {\em Phys. Rev. C} {\bf 94} (2016), no.~3 034316,
  [\href{http://xxx.lanl.gov/abs/1604.07799}{{\tt 1604.07799}}].

\bibitem{Klein:1999qj}
S.~Klein and J.~Nystrand, {\it {Exclusive vector meson production in
  relativistic heavy ion collisions}},  {\em Phys. Rev.} {\bf C60} (1999)
  014903, [\href{http://xxx.lanl.gov/abs/hep-ph/9902259}{{\tt
  hep-ph/9902259}}].

\bibitem{Lewin:1995rx}
J.~Lewin and P.~Smith, {\it {Review of mathematics, numerical factors, and
  corrections for dark matter experiments based on elastic nuclear recoil}},
  {\em Astropart. Phys.} {\bf 6} (1996) 87--112.

\bibitem{Ong:2010gf}
A.~Ong, J.~Berengut, and V.~Flambaum, {\it {The Effect of spin-orbit nuclear
  charge density corrections due to the anomalous magnetic moment on
  halonuclei}},  {\em Phys. Rev. C} {\bf 82} (2010) 014320,
  [\href{http://xxx.lanl.gov/abs/1006.5508}{{\tt 1006.5508}}].

\bibitem{Bertozzi:1972jff}
W.~Bertozzi, J.~Friar, J.~Heisenberg, and J.~Negele, {\it {Contributions of
  neutrons to elastic electron scattering from nuclei}},  {\em Phys. Lett. B}
  {\bf 41} (1972) 408--414.

\bibitem{Xiong:2019umf}
W.~Xiong {\em et~al.}, {\it {A small proton charge radius from an
  electron--proton scattering experiment}},  {\em Nature} {\bf 575} (2019),
  no.~7781 147--150.

\bibitem{Bezginov:2019mdi}
N.~Bezginov, T.~Valdez, M.~Horbatsch, A.~Marsman, A.~Vutha, and E.~Hessels,
  {\it {A measurement of the atomic hydrogen Lamb shift and the proton charge
  radius}},  {\em Science} {\bf 365} (2019), no.~6457 1007--1012.

\bibitem{elementIsotopes}
J.~R. de~Laeter, J.~K. Böhlke, P.~D. Bièvre, H.~Hidaka, H.~S. Peiser,
  K.~J.~R. Rosman, and P.~D.~P. Taylor, {\it Atomic weights of the elements.
  review 2000 (iupac technical report)},  {\em Pure and Applied Chemistry} {\bf
  75} (2003), no.~6 683 -- 800.

\bibitem{Akimov:2018ghi}
{\bf COHERENT} Collaboration, D.~Akimov {\em et~al.}, {\it {COHERENT 2018 at
  the Spallation Neutron Source}},
  \href{http://xxx.lanl.gov/abs/1803.09183}{{\tt 1803.09183}}.

\bibitem{Collar:2019ihs}
J.~I. Collar, A.~R.~L. Kavner, and C.~M. Lewis, {\it {Response of CsI[Na] to
  Nuclear Recoils: Impact on Coherent Elastic Neutrino-Nucleus Scattering
  (CE$\nu$NS)}},  {\em Phys. Rev.} {\bf D100} (2019), no.~3 033003,
  [\href{http://xxx.lanl.gov/abs/1907.04828}{{\tt 1907.04828}}].

\bibitem{HoMeSh:prep}
M.~Hofericher, J.~Menéndez, and A.~Schwenk.
\newblock {In preparation}.

\bibitem{BohrMottelson}
A.~Bohr and B.~R. Mottelson, {\em Nuclear Structure, Vol. I}.
\newblock World Scientific, 1998.

\bibitem{Co:2020gwl}
G.~Co', M.~Anguiano, and A.~Lallena, {\it {Nuclear structure uncertainties in
  coherent elastic neutrino-nucleus scattering}},  {\em JCAP} {\bf 04} (2020)
  044, [\href{http://xxx.lanl.gov/abs/2001.04684}{{\tt 2001.04684}}].

\end{thebibliography}\endgroup

\end{document}